\documentclass[oldversion]{aa}
\usepackage{epsfig}
\def\lsim{\vcenter{\hbox{$<$}\offinterlineskip\hbox{$\sim$}}}
\def\gsim{\vcenter{\hbox{$>$}\offinterlineskip\hbox{$\sim$}}}
\begin{document}
\renewcommand{\topfraction}{1.}
\renewcommand{\bottomfraction}{1.}
\renewcommand{\textfraction}{0.}
\title{Molecules and dust production in the Magellanic Clouds\thanks{Based on
observations collected at the European Southern Observatory, Chile (ESO
N$^{\rm o}$ 75.D-0696 and 77.D-0614).}}
\author{Jacco Th. van Loon\inst{1},
        Martin Cohen\inst{2},
        Joana M. Oliveira\inst{1},
        Mikako Matsuura\inst{3,4},
        Iain McDonald\inst{1},
        Gregory C. Sloan\inst{5},
        Peter R. Wood\inst{6}, and
        Albert A. Zijlstra\inst{7}}
\institute{Astrophysics Group, Lennard-Jones Laboratories, Keele University,
           Staffordshire ST5 5BG, UK
      \and Radio Astronomy Lab, 601 Campbell Hall, University of California at
           Berkeley, Berkeley CA 94720-3411, USA
      \and National Astronomical Observatory of Japan, Osawa 2-21-1, Mitaka,
           Tokyo 181-8588, Japan
      \and Department of Physics and Astronomy, University College London
           Gower Street, London WC1E 6BT, UK
      \and Cornell University, Astronomy Department, Ithaca, NY 14853-6801,
           USA
      \and Research School of Astronomy and Astrophysics, Australian National
           University, Cotter Road, Weston Creek, ACT 2611, Australia
      \and Jodrell Bank Centre for Astrophysics, The University of Manchester,
           Alan Turing Building, Manchester M13 9PL, UK}
\date{Received date; accepted date}
\titlerunning{Molecules and dust production in the Magellanic Clouds}
\authorrunning{van Loon et al.}
\abstract{We present ESO/VLT spectra in the 2.9--4.1 $\mu$m range for a large
sample of infrared stars in the Small Magellanic Cloud (SMC), mainly carbon
stars, massive oxygen-rich Asymptotic Giant Branch (AGB) stars, and red
supergiants. Strong emission from Polycyclic Aromatic Hyrdrocarbons (PAHs) is
detected in the spectrum of the post-AGB object MSX\,SMC\,29. Water ice is
detected in at least one Young Stellar Object, IRAS\,01042$-$7215, for the
first time in the SMC. The strength and shapes of the molecular bands detected
in the evolved stars are compared with similar data for stars in the Large
Magellanic Cloud (LMC). Absorption from acetylene in carbon stars is found to
be equally strong in the SMC as in the LMC, but the LMC stars show stronger
dust emission in their infrared colours and veiling of the molecular bands.
This suggests that a critical link exists in the formation of dust from the
molecular atmosphere in carbon stars which scales with the initial
metallicity. Nucleation seeds based on a secondary element such as titanium or
silicon provide a plausible explanation. In oxygen-rich stars, both the
nucleation seeds and molecular condensates depend on secondary elements (in
particular titanium, silicon, and/or aluminium), which explains the observed
lower molecular abundances and lower dust content in the SMC stars. Emission
from silicon monoxide seen in some oxygen-rich AGB stars and red supergiants
in the SMC suggests that these metal-poor stars are able to drive strong
pulsation shocks through their molecular layers. Data for pulsating dusty AGB
stars and supergiants in the LMC are used to show that pulsation is likely the
critical factor in driving mass loss, as long as dust forms, rather than the
stellar luminosity. Finally, we suggest that the reduced dust production and
consequently slower winds of metal-poor AGB stars and red supergiants are more
likely to result in chemical inhomogeneities and small-scale structure in the
interstellar medium.
}
\maketitle

\section{Introduction}

We owe our existence to the nuclear processing of light elements into heavy
elements inside of stars, and their subsequent dispersal into interstellar
space by way of stellar mass loss. One of the main contributors of carbon and
nitrogen, and arguably the most important ``factory'' of cosmic dust,
Asymptotic Giant Branch (AGB) stars represent the final stages of evolution of
intermediate-mass stars ($M_{\rm ZAMS}\la$1 to $\sim$8 M$_\odot$) during which
they lose up to $\ga$80\% of their mass at rates between $\dot{M}\sim10^{-6}$
to $10^{-4}$ M$_\odot$ yr$^{-1}$. AGB stars thus chemically enrich the local
interstellar medium (ISM) on timescales ranging from less than the dynamical
timescale of a galaxy to as long as the age of the Universe. Massive stars
($M_{\rm ZAMS}\ga$8 M$_\odot$) may become red supergiants (RSGs) and
experience similar dusty mass loss too; this greatly affects the properties of
the progenitor and circumstellar environment of the subsequent core-collapse
supernova, and makes them an important source of dust produced in starbursts
observed at cosmological distances.

Mass loss from red (super)giants happens when strong radial pulsations elevate
the stellar atmosphere to a height where the temperature is sufficiently low,
but the density is still high, for dust formation to occur. Radiation pressure
from these luminous stars ($L\sim5\times10^3-5\times10^5$ L$_\odot$) is
believed to drive away the dust, dragging the gas along with it (e.g., Bowen
1988; Fleischer, Gauger \& Sedlmayr 1992; H\"ofner, Feuchtinger \& Dorfi
1995), although the threshold luminosity for this to occur in oxygen-rich
environments that produce relatively transparent grains (Ferrarotti \& Gail
2006, and references therein) may be uncomfortably high to explain the mass
loss from most oxygen-rich AGB stars (Woitke 2006). At the height of their
mass loss dust-enshrouded AGB stars --- and in extreme cases even RSGs ---
vanish at optical wavelengths but shine brightly in the infrared (IR).

An unsolved problem of red (super)giant mass loss is how the simple molecules
in the stellar photosphere grow into larger assemblies that form the cores for
dust growth (scenarios have been proposed by, e.g., Gail \& Sedlmayr 1988),
and how efficient these processes are at the low metallicity that is
characteristic for the early Universe. Little is known about the molecular
abundances within the dust-formation zone and the dust-to-gas ratio in the
stellar wind. The transformation of some oxygen-rich AGB stars into carbon
stars is known to have extremely important consequences: the molecular and
dust formation chemistry will change from oxygen-dominated to
carbon-dominated, giving rise to a vastly different array of molecules and
dust particles. Hence these two types of AGB star enrich the ISM with very
different material. On the other hand, to solve the opacity problem to drive
oxygen-rich winds, H\"ofner \& Andersen (2007) proposed that even oxygen-rich
giants may form some carbonaceous grains.

Much of what we know about the mass loss from AGB stars and RSGs has been
based on IR observations of dust-enshrouded AGB stars in the Large Magellanic
Cloud (LMC, at 50 kpc) and Small Magellanic Cloud (SMC, at 60 kpc), where
accurate luminosities and mass-loss rates can be obtained (e.g., Wood et al.\
1992; Zijlstra et al.\ 2006; van Loon et al.\ 1999b, 2005a; van Loon 2000;
Matsuura et al.\ 2005, 2006; Marshall et al.\ 2004; Sloan et al.\ 2006;
Groenewegen et al.\ 2007; Lagadec et al.\ 2007). The Magellanic environments
are metal-poor compared to the Sun; their ISM and star clusters (except the
oldest ones) have metallicities of typically [Fe/H]$\sim40$\% in the LMC and
$\sim15$\% in the SMC (e.g., Westerlund 1997; van Loon, Marshall \& Zijlstra
2005).

The above studies suggest that metal-poor carbon stars are surprisingly
abundant in C$_2$ and C$_2$H$_2$, probably because they reach higher C/O
ratios than their Galactic solar-metallicity equivalents. This may be due to
the oxygen-poor photospheres of metal-poor carbon stars, rather than a larger
production of primary carbon. This could have important consequences for the
condensation of carbonaceous dust grains in low-metallicity environments (cf.\
Mattsson et al.\ 2008) --- which might however depend on seeds sensitive to
initial metallicity such as TiC$_2$ (Bernatowicz et al.\ 1991). It is as yet
unknown how the molecular abundances depend on metallicity in oxygen-rich
stars, but here too the nucleation seeds for dust condensation may be
restricted to species such as TiO and ZrO, which are then coated, first by
aluminium-oxides and then by silicates (Vollmer et al.\ 2006; Nittler et al.\
2008); this reliance on elements which are not synthesized inside the stars
themselves (Ti, Al, Si) is consistent with the observed initial-metallicity
dependence of oxygen-rich outflows (van Loon 2006; Marshall et al.\ 2004).

Until now, 3--4 $\mu$m spectra of stars in the SMC were limited to two carbon
stars published by van Loon, Zijlstra \& Groenewegen (1999a), and another
carbon star and one post-AGB object published by Matsuura et al.\ (2005). Here
we present ESO/VLT 3--4 $\mu$m spectra of a large sample of dust-enshrouded
carbon stars, oxygen-rich AGB stars and red supergiants in the SMC, an
analysis of their molecular atmospheres and a comparison with previous LMC
samples (Matsuura et al.\ 2005; van Loon et al.\ 2006). We also present
spectra of two R\,CrB type stars, a post-AGB object showing emission from
Polycyclic Aromatic Hydrocarbons (PAHs), and the first 3--4 $\mu$m spectra of
Young Stellar Objects (YSOs) in the SMC showing water ice absorption and
hydrogen recombination lines.

\section{Observations}

\subsection{Target selection}

The targets were selected from IRAS sources (Loup et al.\ 1997; Groenewegen \&
Blommaert 1998), sources observed with ISO (Groenewegen et al.\ 2000; Cioni et
al.\ 2003), and Spitzer IRS targets (Kraemer et al.\ 2005, 2006; Sloan et al.\
2006; Lagadec et al\ 2007; Sloan et al.\ 2008). These include six stars in the
intermediate-age populous star cluster NGC\,419, of which the four bluest and
least dusty stars would probably not have been selected had they not been part
of the cluster sample of Spitzer targets. Of these 76 targets, 57 were
actually observed by us --- this includes one carbon star and one candidate
post-AGB object published by Matsuura et al.\ (2005). The spectroscopic sample
thus comprises 36 carbon stars, 14 oxygen-rich AGB stars or RSGs, and 7
peculiar sources (two R\,CrB-type stars, two post-AGB objects and three
candidate YSOs). They are listed in Tables 1--3.

\subsection{Spectroscopy}

%
%
\begin{table*}
\caption[]{List of carbon stars, in order of increasing Right Ascension (2MASS
coordinates, in J2000). S[..] identifiers are from Groenewegen \& Blommaert
(1998), ISO-MCMS\,J[..] identifiers are from Cioni et al.\ (2003), and
NGC\,419\,LE[..] identifiers are from Lloyd Evans (1980b). Near-IR magnitudes
are from the 2MASS (JHK$_{\rm s}$) and the spectroscopy acquisition images
(L$_{\rm NB}^\prime$).}
\begin{tabular}{llcccccccc}
\hline\hline
Name                                            &
Alternative(s)                                  &
$\alpha$ ($^{\rm h}$ $^{\rm m}$ $^{\rm s}$)     &
$\delta$ ($^\circ$ $^\prime$ $^{\prime\prime}$) &
Date                                            &
J                                               &
H                                               &
K$_{\rm s}$                                     &
L$^\prime_{\rm NB}$                             &
Note                                            \\
\hline
IRAS\,00271$-$7120 &
S1             &
0 29 19.2      &
$-$71 03 50    &
21/9/05        &
\llap{$>$}17.6 &
\llap{$>$}16.2 &
\llap{1}4.02   &
8.80           &
a              \\
GM\,780        &
               &
0 35 37.3      &
$-$73 09 56    &
19/9/05        &
\llap{1}2.90   &
\llap{1}1.38   &
\llap{1}0.25   &
8.70           &
c              \\
IRAS\,00388$-$7401 &
S25            &
0 40 43.9      &
$-$73 45 24    &
22/9/05        &
\llap{1}5.83   &
\llap{1}3.54   &
\llap{1}1.46   &
9.01           &
a              \\
IRAS\,00393$-$7326 &
S4             &
0 41 14.4      &
$-$73 10 09    &
20/9/05        &
\llap{1}6.26   &
\llap{1}3.97   &
\llap{1}1.97   &
9.40           &
a              \\
MSX\,SMC\,54   &
               &
0 43 05.9      &
$-$73 21 41    &
27/8/06        &
\llap{1}6.60   &
\llap{1}4.50   &
\llap{1}2.57   &
9.32           &
b              \\
MSX\,SMC\,44   &
               &
0 43 39.6      &
$-$73 14 58    &
26/8/06        &
\llap{1}2.57   &
\llap{1}1.15   &
\llap{1}0.03   &
9.06           &
b              \\
MSX\,SMC\,105  &
               &
0 45 02.2      &
$-$72 52 24    &
25/8/06        &
\llap{1}5.31   &
\llap{1}3.00   &
\llap{1}1.19   &
9.45           &
b              \\
MSX\,SMC\,36   &
               &
0 45 54.0      &
$-$73 23 41    &
28/8/06        &
\llap{1}5.07   &
\llap{1}3.23   &
\llap{1}1.60   &
\llap{1}0.54   &
b              \\
IRAS\,00448$-$7332 &
MSX\,SMC\,60, S6 &
0 46 40.4      &
$-$73 16 47    &
21/9/05        &
\llap{1}5.68   &
\llap{1}3.15   &
\llap{1}1.20   &
8.46           &
a,b            \\
MSX\,SMC\,33   &
               &
0 47 05.5      &
$-$73 21 33    &
25/8/06        &
\llap{1}3.45   &
\llap{1}1.70   &
\llap{1}0.32   &
8.27           &
b              \\
IRAS\,00454$-$7257 &
S7             &
0 47 18.9      &
$-$72 40 34    &
20/9/05        &
\llap{$>$}13.1 &
\llap{1}3.59   &
\llap{1}1.65   &
8.01           &
a              \\
               &
               &
               &
               &
26/8/06        &
               &
               &
               &
9.08           &
               \\
IRAS\,00468$-$7341 &
S8             &
0 48 35.7      &
$-$74 10 39    &
21/9/05        &
\llap{$>$}18.4 &
\llap{$>$}16.5 &
\llap{1}3.90   &
9.96           &
a              \\
MSX\,SMC\,66   &
               &
0 48 52.5      &
$-$73 08 57    &
26/8/06        &
\llap{1}4.70   &
\llap{1}2.70   &
\llap{1}1.12   &
8.28           &
b,i            \\
RAW\,631       &
MSX\,SMC\,134  &
0 50 44.4      &
$-$72 37 39    &
26/8/06        &
\llap{1}3.02   &
\llap{1}1.82   &
\llap{1}1.22   &
9.72           &
b,g            \\
MSX\,SMC\,163  &
               &
0 51 00.8      &
$-$72 25 19    &
25/8/06        &
\llap{1}4.74   &
\llap{1}2.70   &
\llap{1}0.99   &
9.03           &
b              \\
ISO-MCMS\,J005149.4$-$731315 &
               &
0 51 49.5      &
$-$73 13 16    &
21/9/05        &
\llap{1}4.26   &
\llap{1}2.56   &
\llap{1}1.17   &
9.10           &
c,e            \\
MSX\,SMC\,125  &
               &
0 51 50.2      &
$-$72 50 50    &
25/8/06        &
\llap{1}3.49   &
\llap{1}1.89   &
\llap{1}0.46   &
8.24           &
g              \\
GM\,106        &
               &
0 53 04.9      &
$-$73 04 10    &
20/9/05        &
\llap{1}4.07   &
\llap{1}2.69   &
\llap{1}1.51   &
9.49           &
c,h            \\
MSX\,SMC\,159  &
               &
0 54 22.3      &
$-$72 43 30    &
27/8/06        &
\llap{1}6.05   &
\llap{1}3.75   &
\llap{1}1.71   &
9.19           &
b              \\
LEGC\,105      &
               &
0 54 46.9      &
$-$73 13 38    &
27/8/06        &
\llap{1}3.27   &
\llap{1}1.80   &
\llap{1}0.73   &
9.50           &
c              \\
ISO-MCMS\,J005450.7$-$730607 &
               &
0 54 50.8      &
$-$73 06 07    &
21/9/05        &
\llap{1}4.15   &
\llap{1}2.44   &
\llap{1}1.00   &
8.88           &
c              \\
ISO-MCMS\,J005454.1$-$730318 &
               &
0 54 54.1      &
$-$73 03 18    &
20/9/05        &
\llap{1}4.35   &
\llap{1}2.36   &
\llap{1}0.79   &
8.40           &
c,j            \\
MSX\,SMC\,209  &
               &
0 56 16.4      &
$-$72 16 41    &
26/8/06        &
\llap{1}3.77   &
\llap{1}2.01   &
\llap{1}0.61   &
9.30           &
b              \\
IRAS\,00554$-$7351 &
S16            &
0 57 04.0      &
$-$73 35 15    &
20/9/05        &
\llap{1}6.39   &
\llap{1}3.56   &
\llap{1}1.41   &
8.82           &
a,c            \\
ISO-MCMS\,J005720.6$-$731245 &
               &
0 57 20.6      &
$-$73 12 46    &
21/9/05        &
\llap{1}3.84   &
\llap{1}2.40   &
\llap{1}1.33   &
9.30           &
c              \\
IRAS\,00557$-$7309 &
S17            &
0 57 27.7      &
$-$72 53 28    &
21/9/05        &
\llap{1}5.93   &
\llap{1}3.68   &
\llap{1}1.86   &
8.00           &
a              \\
MSX\,SMC\,93   &
               &
0 59 23.4      &
$-$73 56 01    &
25/8/06        &
\llap{1}4.42   &
\llap{1}2.71   &
\llap{1}1.41   &
8.78           &
b              \\
NGC\,419\,LE\,16 &
               &
1 08 01.1      &
$-$72 53 17    &
22/9/05        &
\llap{1}3.23   &
\llap{1}1.87   &
\llap{1}0.84   &
9.71           &
c,f            \\
NGC\,419\,IR1  &
               &
1 08 13.0      &
$-$72 52 44    &
22/9/05        &
\llap{1}3.48   &
\llap{1}2.04   &
\llap{1}0.89   &
8.76           &
c,f            \\
NGC\,419\,IR2  &
               &
1 08 17.5      &
$-$72 53 10    &
22/9/05        &
-              &
-              &
-              &
\llap{1}0.17   &
c,f            \\
NGC\,419\,LE\,35 &
               &
1 08 17.5      &
$-$72 53 01    &
22/9/05        &
\llap{1}2.52   &
\llap{1}1.50   &
\llap{1}0.75   &
\llap{1}0.06   &
c,f            \\
NGC\,419\,LE\,27 &
               &
1 08 20.7      &
$-$72 52 52    &
22/9/05        &
\llap{1}2.76   &
\llap{1}1.66   &
\llap{1}1.00   &
\llap{1}0.01   &
c,f            \\
NGC\,419\,LE\,18 &
RAW\,1553      &
1 08 25.0      &
$-$72 52 57    &
22/9/05        &
\llap{1}2.94   &
\llap{1}1.70   &
\llap{1}1.02   &
\llap{1}0.01   &
c,f            \\
GM\,522        &
RAW\,1559      &
1 08 35.4      &
$-$73 26 07    &
25/8/06        &
\llap{1}3.86   &
\llap{1}2.53   &
\llap{1}1.44   &
9.90           &
c              \\
IRAS\,01091$-$7320 &
S23            &
1 10 32.2      &
$-$73 05 04    &
22/9/05        &
\llap{$>$}16.7 &
\llap{1}4.68   &
\llap{1}2.34   &
9.73           &
a              \\
IRAS\,01210$-$7125 &
S30            &
1 22 28.3      &
$-$71 09 26    &
24/7/02        &
\llap{1}4.91   &
\llap{1}2.95   &
\llap{1}1.20   &
(8.1)          &
d              \\
\hline
\vspace{-2mm}
\end{tabular}\\
Notes: a = observed with ISO (Groenewegen et al.\ 2000); b = Spitzer spectrum
(Sloan et al.\ 2006); c = Spitzer spectrum (Lagadec et al.\ 2007); d = L-band
spectrum published in Matsuura et al.\ (2005); e = carbon star RAW\,704
(BMB-B\,70) is at $5^{\prime\prime}$ to SE; f = see Tanab\'e et al.\ (2004)
and van Loon, Marshall \& Zijlstra (2005); g = Spitzer spectrum (Sloan et al.\
in prep.); h = Spitzer spectrum is of PMMR\,52 (see Table 4); i =
IRAS\,00472$-$7325 at $1^\prime$ to SE; j = carbon star RAW\,892 at
$10^{\prime\prime}$ to SE.
\end{table*}

%
%
\begin{table*}
\caption[]{List of dust-enshrouded oxygen-rich AGB stars and red supergiants.
See the description of Table 1.}
\begin{tabular}{llccccccclc}
\hline\hline
Name                                            &
Alternative(s)                                  &
$\alpha$ ($^{\rm h}$ $^{\rm m}$ $^{\rm s}$)     &
$\delta$ ($^\circ$ $^\prime$ $^{\prime\prime}$) &
Date                                            &
J                                               &
H                                               &
K$_{\rm s}$                                     &
L$^\prime_{\rm NB}$                             &
SpT                                             &
Note                                            \\
\hline
HV\,1375       &
MSX\,SMC\,24   &
0 42 52.2      &
$-$73 50 52    &
28/8/06        &
\llap{1}1.30   &
\llap{1}0.38   &
9.83           &
8.64           &
M5             &
b              \\
MSX\,SMC\,18   &
               &
0 46 31.6      &
$-$73 28 46    &
25/8/06        &
\llap{1}2.66   &
\llap{1}1.16   &
\llap{1}0.20   &
8.01           &
               &
b              \\
BFM\,1         &
               &
0 47 19.2      &
$-$72 40 05    &
26/8/06        &
\llap{1}2.66   &
\llap{1}1.65   &
\llap{1}1.03   &
9.72           &
S              &
c              \\
HV\,11262      &
MSX\,SMC\,67   &
0 47 36.9      &
$-$73 04 44    &
25/8/06        &
9.47           &
8.60           &
8.32           &
7.98           &
M0.5           &
b              \\
PMMR\,34       &
MSX\,SMC\,96   &
0 50 06.4      &
$-$73 28 11    &
27/8/06        &
9.59           &
8.76           &
8.54           &
8.45           &
M0.5           &
b              \\
IRAS\,00483$-$7347 &
S9             &
0 50 07.2      &
$-$73 31 25    &
20/9/05        &
\llap{1}1.43   &
9.78           &
8.64           &
6.84           &
M8             &
b              \\
GM\,103        &
IRAS\,00486$-$7308, S10 &
0 50 30.6      &
$-$72 51 30    &
20/9/05        &
\llap{1}0.34   &
9.39           &
8.78           &
8.18           &
M4             &
a,d            \\
PMMR\,41       &
MSX\,SMC\,109  &
0 51 29.7      &
$-$73 10 44    &
27/8/06        &
9.22           &
8.41           &
8.16           &
7.81           &
M0\,I\rlap{ab} &
b              \\
BMB-B\,75      &
               &
0 52 12.9      &
$-$73 08 53    &
27/8/06        &
\llap{1}1.25   &
\llap{1}0.47   &
9.92           &
8.71           &
M              &
b,e            \\
HV\,1652       &
MSX\,SMC\,168  &
0 55 26.8      &
$-$72 35 56    &
28/8/06        &
9.80           &
9.01           &
8.76           &
8.30           &
M0.5           &
b              \\
HV\,11417      &
MSX\,SMC\,149, &
1 00 48.2      &
$-$72 51 02    &
21/9/05        &
9.75           &
8.81           &
8.45           &
7.80           &
M5\,e          &
a,b            \\
               &
IRAS\,00591$-$7307, S18 &
               &
               &
               &
               &
               &
               &
               &
               &
               \\
IRAS\,01066$-$7332 &
S20a           &
1 08 10.3      &
$-$73 15 52    &
25/8/06        &
\llap{1}1.19   &
\llap{1}0.20   &
9.59           &
9.09           &
M8             &
a,b            \\
HV\,12956      &
IRAS\,01074$-$7140, S22 &
1 09 02.3      &
$-$71 24 10    &
22/9/05        &
\llap{1}1.53   &
\llap{1}0.85   &
\llap{1}0.34   &
9.15           &
M5             &
a,f            \\
HV\,2084       &
IRAS\,01082$-$7335, S29 &
1 09 38.2      &
$-$73 20 02    &
28/8/06        &
8.85           &
8.14           &
7.79           &
7.49           &
M2\,I\rlap{a}  &
b              \\
\hline
\vspace{-2mm}
\end{tabular}\\
Notes: a = observed with ISO (Groenewegen et al.\ 2000); b = Spitzer spectrum
(Sloan et al.\ in prep.); c = star of S type (Blanco, Frogel \& McCarthy 1981;
Sloan et al.\ 2008); d = groundbased mid-IR spectrum (Groenewegen et al.\
1995); e = has been confused with a carbon star, a radio continuum
source is at $6^{\prime\prime}$ to W (Filipovi\'c et al.\ 2002); f = Li-rich
(Smith et al.\ 1995).
\end{table*}

%
%
\begin{table*}
\caption[]{List of peculiar objects. See the description of Table 1.}
\begin{tabular}{llcccccccc}
\hline\hline
Name                                            &
Alternative(s)                                  &
$\alpha$ ($^{\rm h}$ $^{\rm m}$ $^{\rm s}$)     &
$\delta$ ($^\circ$ $^\prime$ $^{\prime\prime}$) &
Date                                            &
J                                               &
H                                               &
K$_{\rm s}$                                     &
L$^\prime_{\rm NB}$                             &
Note                                            \\
\hline
\multicolumn{10}{l}{\it R\,CrB-type stars} \\
MSX\,SMC\,14   &
               &
0 46 16.3      &
$-$74 11 14    &
27/8/06        &
\llap{$>$}17.4 &
\llap{$>$}15.8 &
\llap{1}3.97   &
9.79           &
c              \\
MSX\,SMC\,155  &
               &
0 57 18.2      &
$-$72 42 35    &
25/8/06        &
\llap{1}3.88   &
\llap{1}2.81   &
\llap{1}1.73   &
8.51           &
c              \\
\multicolumn{10}{l}{\it Post-AGB (C) objects} \\
MSX\,SMC\,29   &
               &
0 36 46.3      &
$-$73 31 35    &
26/8/06        &
\llap{$>$}16.6 &
\llap{1}5.75   &
\llap{1}3.50   &
\llap{1}0.47   &
d              \\
IRAS\,00350$-$7436 &
S2             &
0 36 59.6      &
$-$74 19 50    &
24/7/02        &
\llap{1}1.32   &
\llap{1}0.16   &
9.13           &
(7.7)          &
e              \\
\multicolumn{10}{l}{\it Candidate Young Stellar Objects} \\
MSX\,SMC\,79   &
               &
0 48 39.7      &
$-$73 25 01    &
27/8/06        &
\llap{1}6.57   &
\llap{1}5.54   &
\llap{1}4.34   &
\llap{1}0.82   &
b              \\
IRAS\,01039$-$7305 &
MSX\,SMC\,180, DEM\,S129 &
1 05 30.3      &
$-$72 49 54    &
28/8/06        &
\llap{1}4.86   &
\llap{1}3.24   &
\llap{1}1.69   &
8.57           &
b              \\
IRAS\,01042$-$7215 &
S28            &
1 05 49.3      &
$-$71 59 49    &
26/8/06        &
\llap{1}6.52   &
\llap{1}5.15   &
\llap{1}3.29   &
9.24           &
a,f            \\
\hline
\vspace{-2mm}
\end{tabular}\\
Notes: a = observed with ISO (Groenewegen et al.\ 2000); b = Spitzer spectrum
(Sloan et al.\ in prep.); c = Spitzer spectrum (Kraemer et al.\ 2005); d =
Spitzer spectrum (Kraemer et al.\ 2006); e = L-band spectrum published by
Matsuura et al.\ (2005); f = Spitzer IRS and MIPS-SED cycle 5 target (P.I.\
Sloan).
\end{table*}

%
%
\begin{table*}
\caption[]{Serendipitous objects, listed in order of increasing Right
Ascension (2MASS coordinates, in J2000). Near-IR magnitudes are from the 2MASS
(JHK$_{\rm s}$) and the spectroscopy acquisition images (L$^\prime_{\rm
NB}$).}
\begin{tabular}{lccccccclc}
\hline\hline
Name                                            &
$\alpha$ ($^{\rm h}$ $^{\rm m}$ $^{\rm s}$)     &
$\delta$ ($^\circ$ $^\prime$ $^{\prime\prime}$) &
Date                                            &
J                                               &
H                                               &
K$_{\rm s}$                                     &
L$^\prime_{\rm NB}$                             &
Type                                            &
\llap{N}ote                                     \\
\hline
2MASS\,J00472001$-$7240350 &
0 47 20.0    &
$-$72 40 35  &
20/9/05      &
\llap{1}1.43 &
\llap{1}0.52 &
\llap{1}0.30 &
\llap{1}0.01 &
O-rich       &
             \\
2MASS\,J00473917$-$7304115 &
0 47 39.2    &
$-$73 04 12  &
25/8/06      &
\llap{1}4.58 &
\llap{1}2.97 &
\llap{1}1.56 &
9.55         &
C star       &
             \\
MSX\,SMC\,79\,B &
0 48 41.0    &
$-$73 25 03  &
27/8/06      &
-            &
-            &
-            &
\llap{1}1.80 &
YSO?         &
             \\
RAW\,616     &
0 50 22.9    &
$-$72 51 41  &
20/9/05      &
\llap{1}2.91 &
\llap{1}1.88 &
\llap{1}1.38 &
\llap{1}0.53 &
Post-AGB (C) object\rlap{?} &
             \\
RAW\,639     &
0 50 50.6    &
$-$72 37 19  &
26/8/06      &
\llap{1}2.87 &
\llap{1}1.74 &
\llap{1}1.19 &
\llap{1}0.46 &
C star       &
             \\
2MASS\,J00513146$-$7310513 &
0 51 31.5    &
$-$73 10 51  &
27/8/06      &
\llap{1}0.85 &
\llap{1}0.10 &
9.86         &
9.59         &
O-rich       &
             \\
PMMR\,52     &
0 53 09.1    &
$-$73 04 04  &
20/9/05      &
9.26         &
8.45         &
8.22         &
7.90         &
K5           &
a            \\
2MASS\,J00544039$-$7313407 &
0 54 40.4    &
$-$73 13 41  &
27/8/06      &
\llap{1}1.77 &
\llap{1}0.91 &
\llap{1}0.60 &
\llap{1}0.20 &
O-rich       &
             \\
PMMR\,79 ?   &
0 55 36.6    &
$-$72 36 24  &
28/8/06      &
9.88         &
9.12         &
8.91         &
8.54         &
K5           &
             \\
2MASS\,J00565642$-$7335301 &
0 56 56.4    &
$-$73 35 30  &
20/9/05      &
\llap{1}4.94 &
\llap{1}3.22 &
\llap{1}1.66 &
\llap{1}0.32 &
C star       &
             \\
RAW\,1030    &
0 57 11.6    &
$-$73 13 15  &
21/9/05      &
\llap{1}2.77 &
\llap{1}1.81 &
\llap{1}1.37 &
\llap{1}0.89 &
C star?      &
             \\
PMMR\,116    &
1 00 54.2    &
$-$72 51 37  &
21/9/05      &
9.52         &
8.92         &
8.68         &
8.58         &
K6.5         &
             \\
\hline
\vspace{-2mm}
\end{tabular}\\
Notes: a = Spitzer spectrum (Lagadec et al.\ 2007).
\end{table*}

The ISAAC instrument on the European Southern Observatory (ESO) Very Large
Telescope (VLT) ``Antu'' at Paranal, Chile, was used on the nights of 19 to 22
September 2005 and 25 to 28 August 2006 to obtain low-resolution long-slit
spectra between 2.9 and 4.1 $\mu$m. At a slit width of $2^{\prime\prime}$ the
spectral resolving power was essentially determined by the seeing and was
$200{\la}R\la400$. The smallest attained resolution element is sampled by
three pixels on the Aladdin array.

The spectra were obtained by chopping and nodding with a throw of
$10^{\prime\prime}$ to cancel the high background at these thermal-IR
wavelengths, and jittering within $2^{\prime\prime}$ to remove the effects of
bad pixels. This produced three spectra on the final combined frame, with the
central spectrum twice as bright as the other (inverse) spectra. Exposure
times varied between 12 and 60 minutes. After removing cosmic ray hits and
correcting the spectra for their small inclination with respect to the array,
all spectra were extracted using an optimal extraction algorithm. The
extracted spectra were then aligned and combined. Wavelengths calibration was
performed with respect to the many telluric absorption lines in the spectra;
the spectra are sampled on a grid of 14-\AA\ elements.

The six targets in the cluster NGC\,419 were observed pairwise simultaneously,
placing both on the slit. The chop/nod throw was enlarged to respectively
$13^{\prime\prime}$ for NGC\,419\,IR2 and NGC\,419\,LE\,35, which are
$8^{\prime\prime}$ apart, and $15^{\prime\prime}$ for NGC\,419\,LE\,27 and
NGC\,419\,LE\,18, which are $20^{\prime\prime}$ apart. In the latter
observation, also light from NGC\,419\,IR1 was caught in the slit, and its
spectrum is combined with that obtained during its dedicated observation
through an exposure time-weighted average.

Targets MSX\,SMC\,79 and PMMR\,41 were observed using chop/nod throws of
$12^{\prime\prime}$ and $15^{\prime\prime}$, respectively, to facilitate the
simultaneous observation of an additional star on the slit. One star,
IRAS\,00454$-$7257 was observed in 2005, and again in 2006 when it could be
placed on the slit and observed simultaneously with another nearby target.

To cancel the telluric absorption and correct for the overall spectral
response in the target stars' spectra, a variety of telluric standard stars
were observed. These were generally main sequence B stars, but we accidentally
also observed two B-type emission-line stars (Appendix A). The imprints left
by the hydrogen absorption lines in the quotient spectra of target and
standard star are at a level of a few per cent. An example of the telluric
absorption affecting the spectra before correction is given in van Loon et
al.\ 2006).

Residuals from telluric line removal and/or bad pixels were suppressed by
applying a running boxcar filter with a width of 100 pixels (0.14 $\mu$m),
replacing data values more deviant than three times the median absolute
deviation as well as more than 10\% of the mid-spectrum intensity level. The
result was inspected visually, and higher thresholds were set in some cases to
prevent sharp absorption features or emission lines being affected by this
procedure.

The observing conditions varied between fair and poor; although the humidity
was quite stable around 30\% and 10\% during the 2005 and 2006 runs,
respectively, the optical seeing had the tendency of increasing from
$\sim1^{\prime\prime}$ at the start of the night to $\sim1.5^{\prime\prime}$
in the 2005 run, and to over $2.5^{\prime\prime}$ in the 2006 run (the IR
seeing was better). Interruptions occurred due to cloud cover, especially in
the 2005 run when clouds would form suddenly upon sunset.

\subsection{Photometry}

IR photometry is compiled for all targets and for the two objects observed by
Matsuura et al.\ (2005), as well as for the serendipitous objects that were
placed on the slit whilst observing one of the primary targets (Table 4).

JHK$_{\rm s}$-band photometry is available from the 2-Micron All-Sky Survey
(2MASS: Skrutskie et al.\ 2006), but some of them are too faint in the J-band
to have been detected in the 2MASS. L$^\prime_{\rm NB}$-band ($\lambda_{\rm
centre}=3.80$ $\mu$m, $\Delta\lambda=0.06$ $\mu$m) photometry is measured from
the spectroscopy acquisition images, calibrated against the acquisition images
of the telluric standard stars and generally accurate to within $\sim0.1$ mag.
Photometry for NGC\,419 is also available from van Loon, Marshall \& Zijlstra
(2005) --- who first detected NGC\,419\,IR2 in the K-band (cf.\ Groenewegen et
al.\ 2007) --- and Tanab\'e et al.\ (2004).

For IRAS\,01210$-$7125 and IRAS\,00350$-$7436, the spectra of which were
published by Matsuura et al.\ (2005), we measured (broad-band) L-band
magnitudes from their acquisition images.

Van Loon et al.\ (2006) and Whitelock et al.\ (2003) published photometry for
sources in the LMC with which we compare our new data, but note that the
L-band filter used by Whitelock et al.\ (2003) differs from the 3.80-$\mu$m
narrowband filter used by us.

\section{Results}

\subsection{Global properties of the sample}

The following subsections describe the spectroscopic identification of the
various types of objects; the results are used already here to describe their
photometric properties.

%
%
\begin{figure}
\centerline{\psfig{figure=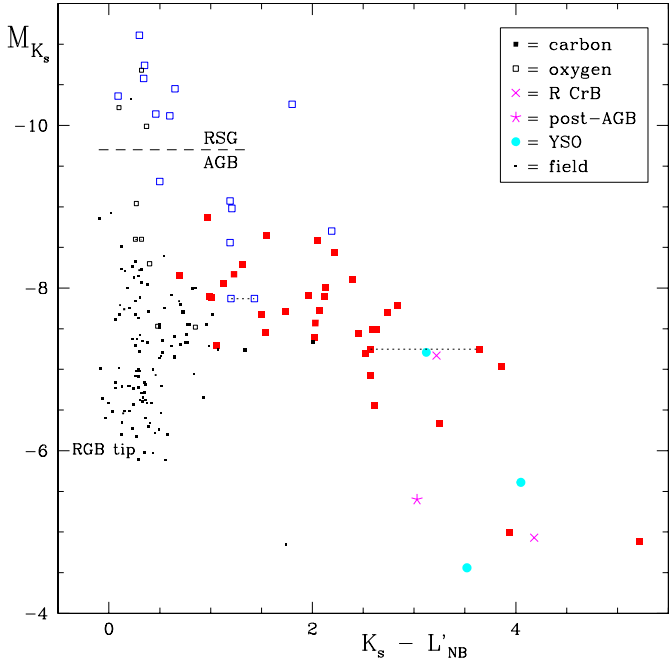,width=88mm}}
\caption[]{2MASS-ISAAC colour-magnitude diagram of targets and field stars,
except the objects in NGC\,419 (see Fig.\ 2). Serendipitously observed stars
are indicated with smaller squares (in black) than the primary targets (in
blue and red).}
\end{figure}

All our targets have K$_{\rm s}$-band and 3.80-$\mu$m photometry; the
corresponding colour-magnitude diagram is displayed in Fig.\ 1, for all but
the NGC\,419 stars which are displayed in Fig.\ 2 instead. The acquisition
images sample stars brighter than the tip of the RGB. Stars that were not
targets for our spectroscopy (``field'') lie on the main AGB sequence ($K_{\rm
s}-L^\prime_{\rm NB}$ $\lsim$ 1 mag and $M_{\rm Ks}$ $\gsim-9$ mag); these
include the many optically bright carbon stars known in the SMC (e.g.,
Rebeirot, Azzopardi \& Westerlund 1993), that have little dust and are
therefore less interesting for the purpose of this study and often rather
faint for 3--4 $\mu$m spectroscopy. Our targets are brighter in L$^\prime_{\rm
NB}$.

There is a clear gap in K$_{\rm s}$-band magnitude between the brightest AGB
stars (which tend to be oxygen-rich) and the RSGs. This may correspond to the
realm of the super-AGB stars, with masses around 8 M$_\odot$ or so, which are
found to become less cool and dusty than AGB stars or slightly more massive
RSGs (van Loon et al.\ 2005a,b). Photometric variability and overluminosity as
a result of Hot Bottom Burning (HBB; Boothroyd \& Sackmann 1992) could cause
an AGB star to be classified as a RSG --- HV\,11417 may be an example of this,
as it has a large K-band amplitude of $\Delta K=1.3$ mag which is uncommon for
stars as luminous as RSGs, and the 2MASS values correspond to near-maximum
light.

Carbon stars start digressing from the vertical AGB sequence around $M_{\rm
Ks}\simeq-7.5$ mag, towards redder colours. They quickly become dusty; stars
redder than the vertical AGB sequence by $\Delta (K_{\rm s}-L^\prime_{\rm
NB})\sim0.7$ mag are almost invariably spectroscopic targets. Oxygen-rich AGB
stars bright enough for our spectroscopic programme are generally brighter
(and presumably more massive) than the carbon stars. Although not a carbon
star, BFM\,1 sits at a locus in the colour-magnitude diagram that is otherwise
dominated by carbon stars. This star is classified as an S-type star with a
carbon-to-oxygen ratio nearly 1 by Blanco, Frogel \& McCarthy (1981), which
was confirmed by the detection of LaO in its optical spectrum by Sloan et al.\
(2008). It is a Mira-type variable (with a 394 days optical period) and
$L^\prime_{\rm NB}=9.60$ and 9.83 mag in 2005 and 2006, respectively.

Rare phases of extreme dust obscuration are seen in the form of a few carbon
stars with $K_{\rm s}-L^\prime_{\rm NB}>3$ mag. The fact that no such obscured
oxygen-rich AGB stars are present in our sample (whereas many Galactic
examples are known; e.g., Jones \& Merrill 1976) may be due to the following
effects: (1) oxygen-rich dust is more transparent at near-IR wavelengths,
especially if the iron content is low (Woitke 2006) which may be the case in
the metal-poor SMC, which together with the generally low dust content in the
SMC renders these stars less obscured; (2) more luminous AGB stars have more
extended and hence more diluted circumstellar envelopes (the optical depth
scales roughly as $\tau \propto r^{-1} \propto L^{-0.5}$, where $r$ is the
inner radius of the dust envelope) --- in this regard, IRAS\,00483$-$7347 is
an extreme example of a dust-enshrouded RSG; (3) the initial mass function
(IMF) and rapid evolution of relatively massive stars conspire to make it
difficult to catch such a star in the most dust-enshrouded phase, especially
in a small galaxy such as the SMC; (4) carbon stars form at lower masses in
the Magellanic Clouds than in the Milky Way (e.g., Marigo, Girardi \& Bressan
1999), thereby enlarging the population of carbon stars at the expense of
low-mass oxygen-rich AGB stars.

The two R\,CrB-type stars and post-AGB object MSX\,SMC\,29 are amongst the
reddest objects, and fainter than the bulk of the carbon stars. The reason for
the faintness may in part be due to their hotter stellar photospheric
temperatures (hence larger IR bolometric corrections), not just dust
obscuration. In that regard, candidate post-AGB object IRAS\,00350$-$7436
(Whitelock et al.\ 1989; Matsuura et al.\ 2005) is rather bright in K.

Three very red objects show the 3-$\mu$m water ice feature and/or hydrogen
emission lines, which suggest these are hot stars embedded in thick, cold dust
cocoons and most likely YSOs. Although some Galactic AGB/post-AGB stars are
known to exhibit the water ice feature (e.g., Smith, Sellgren \& Tokunaga
1988), this is not commonly observed; the location of these three objects in
obvious star-forming regions makes an identification as YSOs more plausible.
One field star, 2MASS\,J00462938$-$7315530 is quite red ($J-K_{\rm s}=2.0$
mag, $K_{\rm s}-L^\prime_{\rm NB}=1.7$ mag) and very faint. It is seen in the
direction of Dark Nebula \#5 identified by Hodge (1974), and it might well be
a star behind this dust cloud or a YSO within it.

\subsubsection{Dusty stars in the populous cluster NGC\,419}

%
%
\begin{figure}
\centerline{\psfig{figure=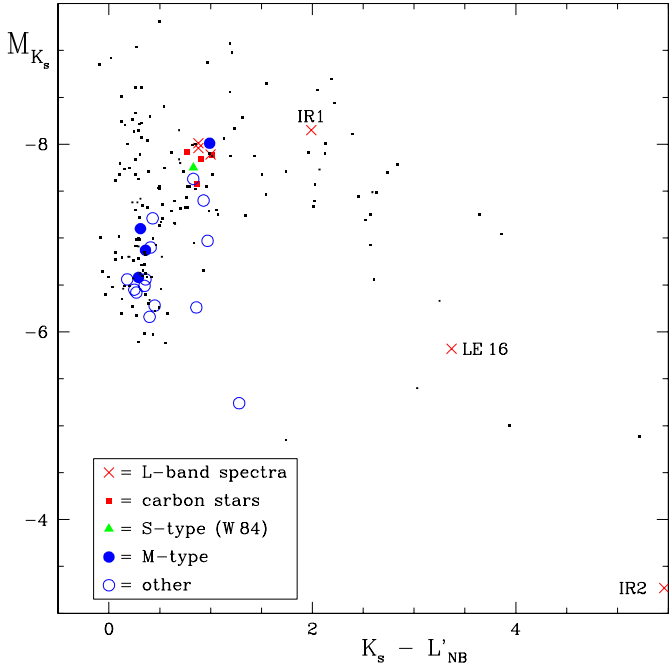,width=88mm}}
\caption[]{SOFI-ISAAC colour-magnitude diagram for the intermediate-age
populous cluster NGC\,419. The stars from Fig.\ 1 (excluding the R\,CrB and
YSO objects and field stars) are plotted as small dots for reference. All 6
targets for which L-band spectra were obtained (crosses) are carbon stars.}
\end{figure}

The stars in NGC\,419 follow the trends outlined above (Fig.\ 2). All stars
for which we obtained L-band spectra are carbon stars, and they are indeed
where IR-bright carbon stars are expected.

NGC\,419\,IR2 (Tanab\'e et al.\ 1999; van Loon et al.\ 2005b) is the most
extremely dust-enshrouded carbon star known in the SMC, but two more carbon
stars in NGC\,419, LE\,16 (Lloyd Evans 1980a,b) and IR1 (Tanab\'e et al.\
1997; van Loon et al.\ 2005b) are also very dusty. LE\,16 is a large-amplitude
variable: the K$_{\rm s}$-band magnitude from the 2MASS is 10.84 mag whilst it
is 13.08 mag on our SOFI images (van Loon et al.\ 2005b). IR2 was detected in
the near-IR only on our SOFI images ($K_{\rm s}=15.63$ mag) and by Groenewegen
et al.\ (2007).

The three other carbon stars for which we obtained L-band spectra are not very
red, and occupy the locus in the colour-magnitude diagram where also an M-type
(Lloyd Evans 1983) and an S-type star (W\,84, Walker 1972) are known to exist
in NGC\,419. Three further confirmed M-type stars (Lloyd Evans 1983) in
NGC\,419 occupy the lower portion of the vertical AGB sequence. A few other
stars in NGC\,419 are clearly redder than the main AGB sequence, but fainter
than the red carbon star branch. It is unclear what causes their red colours,
but interstellar extinction is a possibility.

\subsection{Spectra of carbon stars}

%
%
\begin{figure*}
\centerline{\psfig{figure=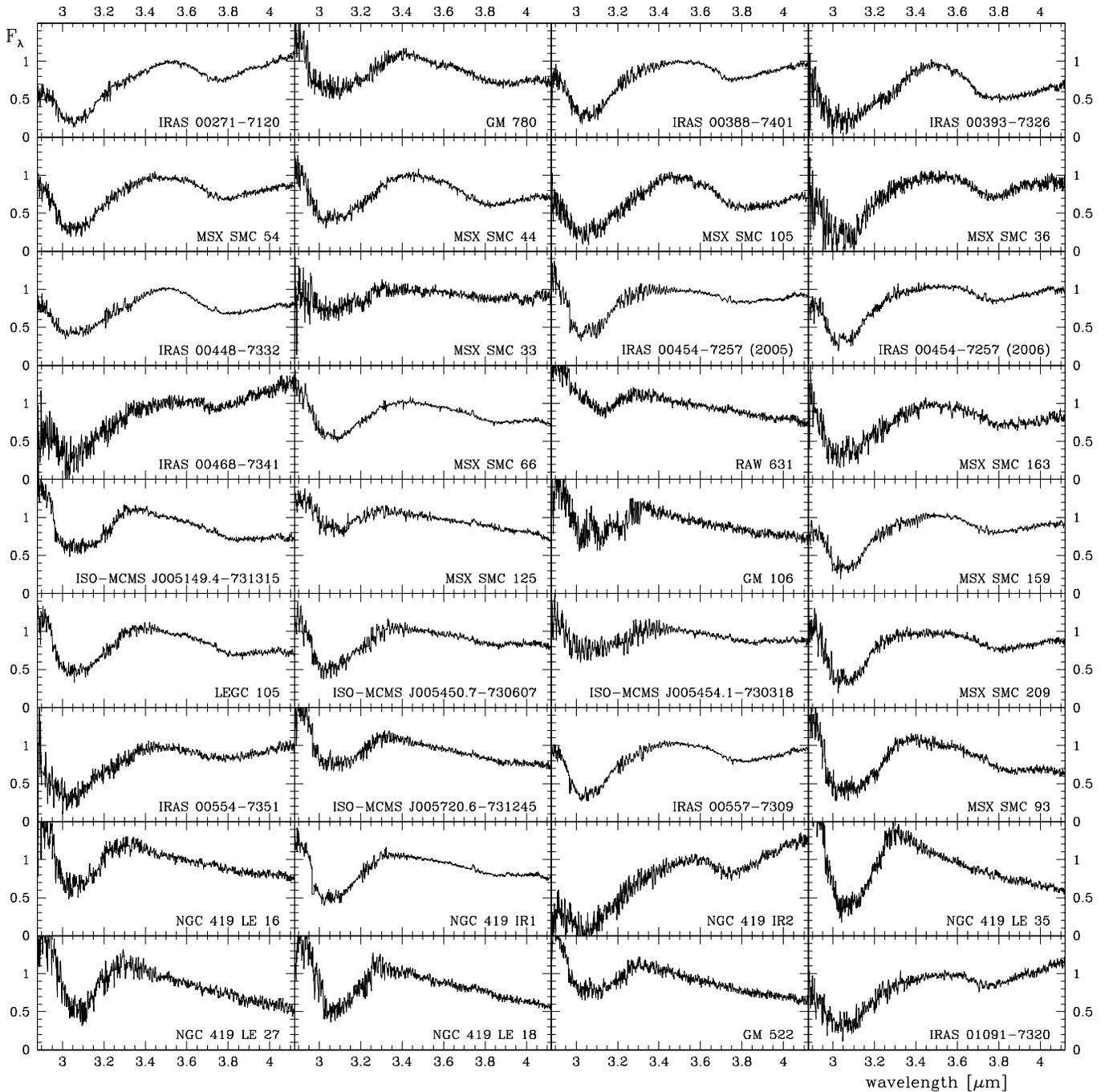,width=180mm}}
\caption[]{VLT/ISAAC 3--4 $\mu$m spectra of IR-bright carbon stars in the
SMC. The most conspicuous features are the broad absorption bands of acetylene
(C$_2$H$_2$) and hydrogen cyanide (HCN) around 3.1 $\mu$m and often also
around 3.8 $\mu$m.}
\end{figure*}

The carbon stars are listed in Table 1, and their spectra are displayed in
Fig.\ 3 (normalised to a flux density of 1 at 3.5 $\mu$m). They all show the
prominent absorption band around 3.1 $\mu$m due to the toxic gases acetylene
(C$_2$H$_2$) and hydrogen cyanide (HCN), and in most cases also the 3.8 $\mu$m
acetylene band. The much sharper 3.57 $\mu$m HCN band is visible in some
stars, but only weakly (e.g., IRAS\,00393$-$7326). The band strengths and
shapes and continuum slopes are discussed in Section 4.1.

GM\,106 was observed during poor conditions, and its 3.1-$\mu$m band is
clearly affected by the weak signal below 3.4 $\mu$m. IRAS\,00393$-$7326 and
IRAS\,00554$-$7351 were the first SMC stars ever for which an L-band spectrum
was published (van Loon et al.\ 1999a), but our new spectra are of a superior
quality, resolution and spectral coverage.

\subsection{Spectra of R\,CrB-type stars}

%
%
\begin{figure}
\centerline{\psfig{figure=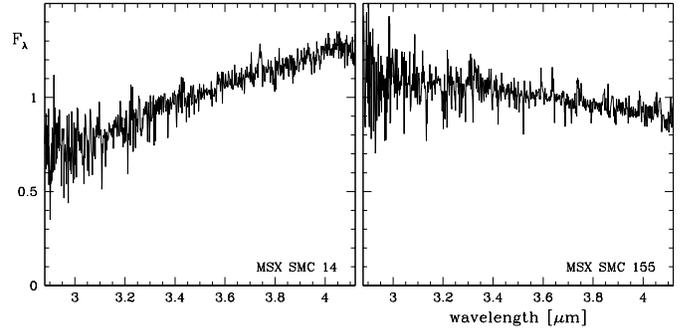,width=88mm}}
\caption[]{VLT/ISAAC 3--4 $\mu$m spectra of two R\,CrB type stars in the SMC,
displaying a featureless dust emission continuum.}
\end{figure}

Two R\,CrB-type stars were observed with the Spitzer IRS (Kraemer et al.\
2005); they are listed in Table 3 and their L-band spectra are displayed in
Fig.\ 4. These are carbon-rich objects which are dominated in this wavelength
region by a dust emission continuum (e.g., Lambert et al.\ 2001). It is clear
both from the photometry and the spectra that MSX\,SMC\,14 is dustier than
MSX\,SMC\,155 (cf.\ Kraemer et al.\ 2005). An optical spectrum of
MSX\,SMC\,155 reveals it to be a rather cool carbon star, suggesting that it
is still near the AGB (Kraemer et al.\ 2005).

\subsection{Spectra of post-AGB objects}

%
%
\begin{figure}
\centerline{\psfig{figure=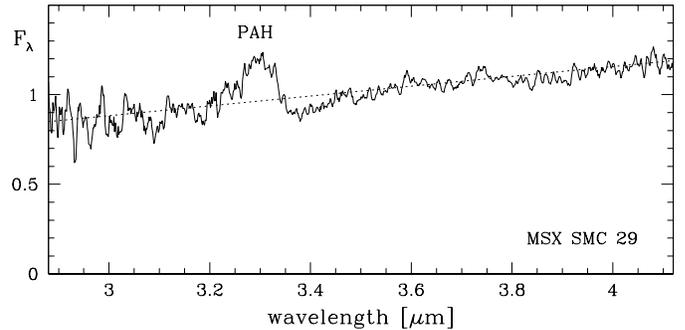,width=88mm}}
\caption[]{VLT/ISAAC 3--4 $\mu$m spectrum of a post-AGB object in the SMC,
displaying a dust emission continuum as well as a prominent emission band at
3.28 $\mu$m attributed to Polycyclic Aromatic Hydrocarbons (PAHs).}
\end{figure}

A post-AGB object observed with the Spitzer IRS (Kraemer et al.\ 2006) is
listed in Table 3; its L-band spectrum is displayed in Fig.\ 5. This object
exhibits a prominent emission band around 3.28 $\mu$m, which is attributed to
PAHs. It also shows a hint of absorption at 3.4 $\mu$m, which we discuss in
Section 4.1.4. Another example of such an object, the L-band spectrum of
IRAS\,00350$-$7436 was published by Matsuura et al.\ (2005); it is listed in
Table 3 for completeness. The presence of PAHs suggests that these post-AGB
objects descended from carbon stars.

\subsection{Spectra of oxygen-rich AGB stars and RSGs}

%
%
\begin{figure}
\centerline{\psfig{figure=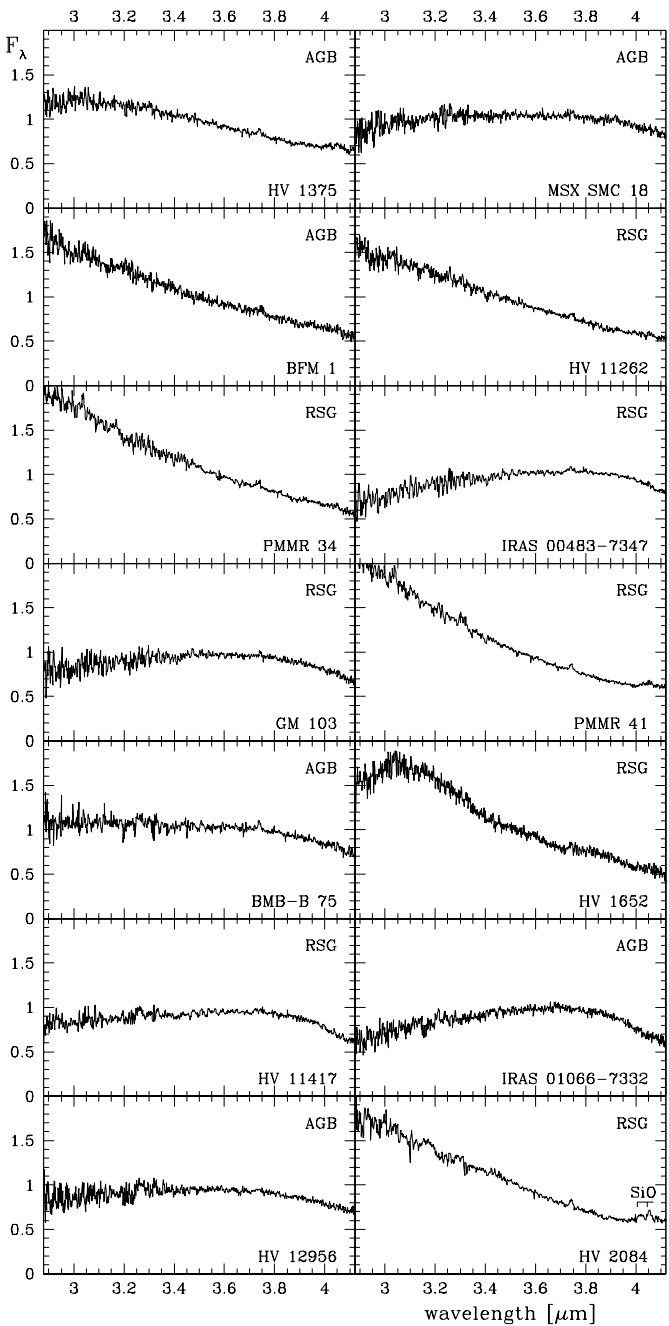,width=88mm}}
\caption[]{VLT/ISAAC 3--4 $\mu$m spectra of oxygen-rich AGB stars and red
supergiants in the SMC. The spectral curvature is determined mostly by the
strength of dust emission and water vapour absorption; in some stars the
silicon monoxide (SiO) band around 4 $\mu$m is seen, either in absorption or
in emission.}
\end{figure}

A collection of oxygen-rich stars were observed (Table 2; Fig.\ 6). As
discussed in Section 3.1 these separate out into AGB stars and RSGs on the
basis of their K-band luminosity. Although there are no clear-cut differences
in spectral appearance between these two classes, RSGs do seem to show the
first overtone SiO band at 4 $\mu$m more often in emission. The best examples
of the latter are also the most luminous in the sample, viz.\ PMMR\,41 and
HV\,2084. The SiO band is discussed in more detail in Section 4.2.

The spectrum of the S-type star BFM\,1 is rather featureless and not very red,
indicating little molecular absorption or dust emission. This is commensurate
with its not very red colours.

BMB-B\,75 is listed in Simbad and recent papers as a carbon star, but it was
discovered as an oxygen-rich AGB star of spectral type M6 (Blanco, Blanco \&
McCarthy 1980). The latter is consistent with our spectrum, with the Spitzer
spectrum showing silicate emission (Sloan et al.\ in prep.), and with the long
pulsation period of 1453 days (Cioni et al.\ 2003).

HV\,1652 shows strong absorption shortward of 3 $\mu$m, possibly water vapour.
There is also a broad depression running from about 3.4 until 3.75 $\mu$m ---
possibly due to OH (Wallace \& Hinkle 2002).

\subsection{Spectra of candidate Young Stellar Objects}

%
%
\begin{figure}
\centerline{\psfig{figure=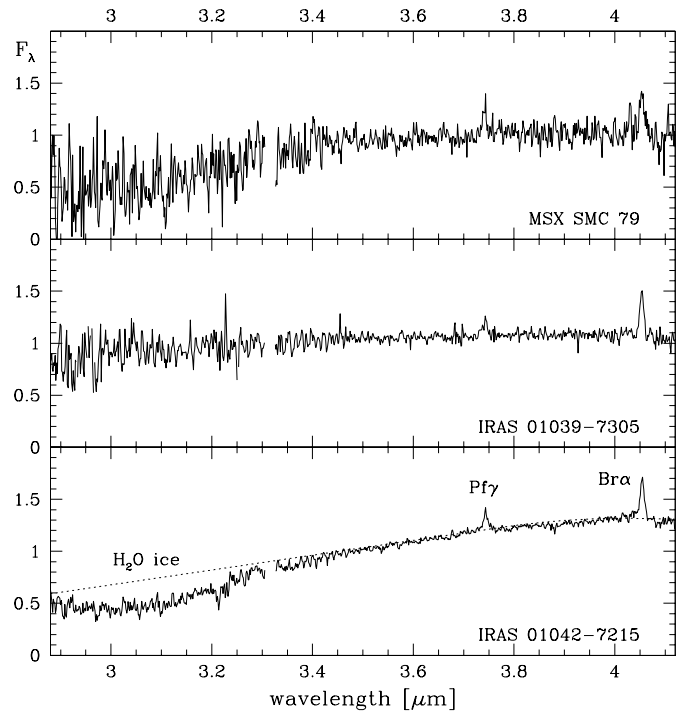,width=88mm}}
\caption[]{VLT/ISAAC 3--4 $\mu$m spectra of YSO candidates in the SMC. All
three show hydrogen line emission, and water ice absorption around 3.1 $\mu$m
is prominent in the reddest of them, IRAS\,01042$-$7215 (the dotted curve is a
$2^{\rm nd}$-order polynomial fit to the continuum beyond 3.6 $\mu$m).}
\end{figure}

Three of the oxygen-rich objects show hydrogen lines in emission, and at least
in one case (IRAS\,01042$-$7215, the reddest of the three) absorption in the
3-$\mu$m band of water ice is prominent (Fig.\ 7; Table 3). MSX\,SMC\,79 is
situated near the young star cluster Bruck\,48. These are good candidates for
being massive embedded Young Stellar Objects (YSOs), and the first such
examples in the SMC for which L-band spectra are presented.

It is interesting to note that in contrast to the first massive YSO studied in
this way in the LMC, IRAS\,05328$-$6827 (van Loon et al.\ 2005c), there is
very little trace of methanol ice (CH$_3$OH) in the spectrum of
IRAS\,01042$-$7215 whilst the water ice band is stronger than in the LMC
object, viz.\ $N({\rm H}_2{\rm O})\sim4\times10^{17}$ cm$^{-2}$. This might
reflect a lower efficiency of ice processing as a result of a reduced total
grain surface area and/or a reduced interstellar abundance of carbon monoxide.
Oliveira et al.\ (2008) present a much more detailed study of the chemistry in
massive YSOs in the SMC, and Oliveira et al.\ (2006) presented more LMC
examples.

\subsection{Serendipitous spectra}

%
%
\begin{figure}
\centerline{\psfig{figure=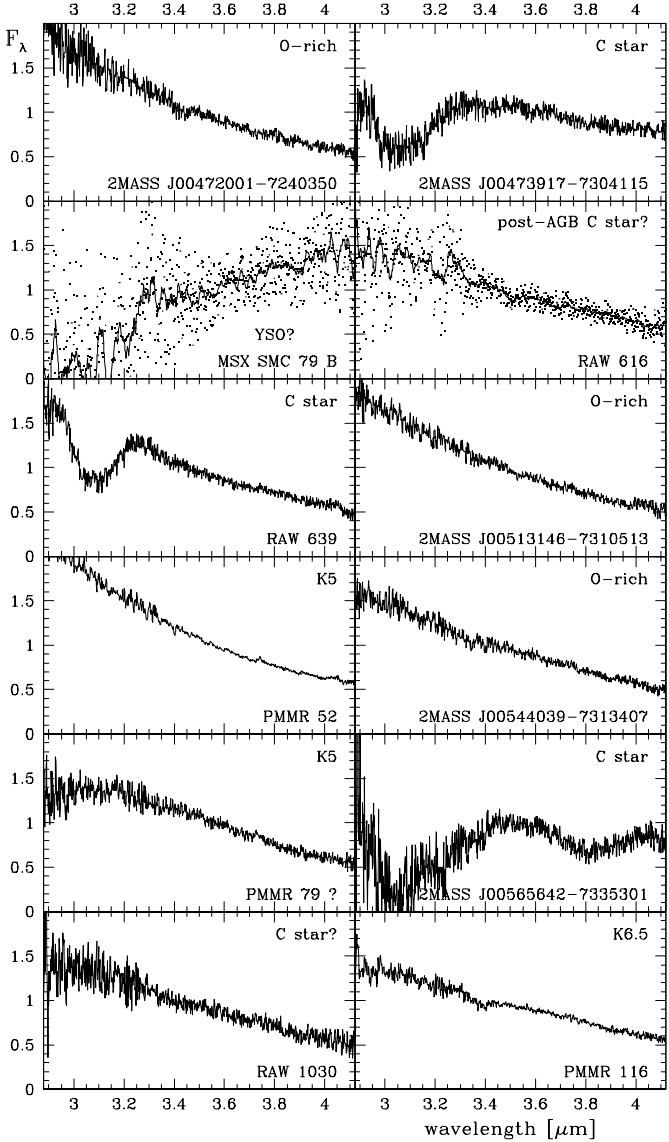,width=88mm}}
\caption[]{VLT/ISAAC 3--4 $\mu$m spectra of stars in the SMC that were
observed simultaneously with primary targets. These objects comprise both
carbon stars and oxygen-rich giants, and possibly an embedded YSO
(MSX\,SMC\,79\,B) and a post-AGB object (RAW\,616) --- the latter two were
heavily smoothed but the original spectral data are also plotted (as dots).}
\end{figure}

Red and/or IR-bright objects were sometimes identified and placed on the slit
to be observed simultaneously with a primary target (Table 4 and Fig.\ 8).

Three are carbon stars: 2MASS\,J00473917$-$7304115, RAW\,639 and
2MASS\,J00565642$-$7335301. Whereas the first two are fairly warm the other is
cool and dusty and shows strong absorption in the 3.8-$\mu$m band of
acetylene. Two more stars are believed to be carbon stars, RAW\,616 and
RAW\,1030, but their spectra show no trace of the 3-$\mu$m band. Veiling by
dust emission cannot be the explanation in this case as these two stars are
not very dusty at all. Late-R-type carbon stars also do not always show the
3-$\mu$m band (e.g., Le Bertre et al.\ 2005). After heavily smoothing the
spectrum, RAW\,616 shows a hint of the 3.28-$\mu$m PAH band in emission, which
means that it might be a carbon star in its post-AGB phase. This is rather
unlikely to have been discovered serendipitously, and it is possible that the
PAH emission arises from the surrounding sky background, as PAH emission is
common in the Magellanic Cloud ISM.

The O-rich stars are rather unremarkable except for the luminous object
PMMR\,52, of which a Spitzer spectrum was taken by Lagadec et al.\ (2007).

The most intriguing serendipitously observed object is MSX\,SMC\,79\,B. The
signal in the spectrum is weak (Fig.\ 8), but it clearly displays an extremely
red continuum, probably from dust emission, and possibly water ice absorption
around 3 $\mu$m and/or PAH emission around 3.28 $\mu$m. Near to YSO candidate
MSX\,SMC\,79 itself, this object too may be an embedded YSO.

\section{Analysis}

\subsection{Carbon chemistry}

%
%
\begin{figure}
\centerline{\psfig{figure=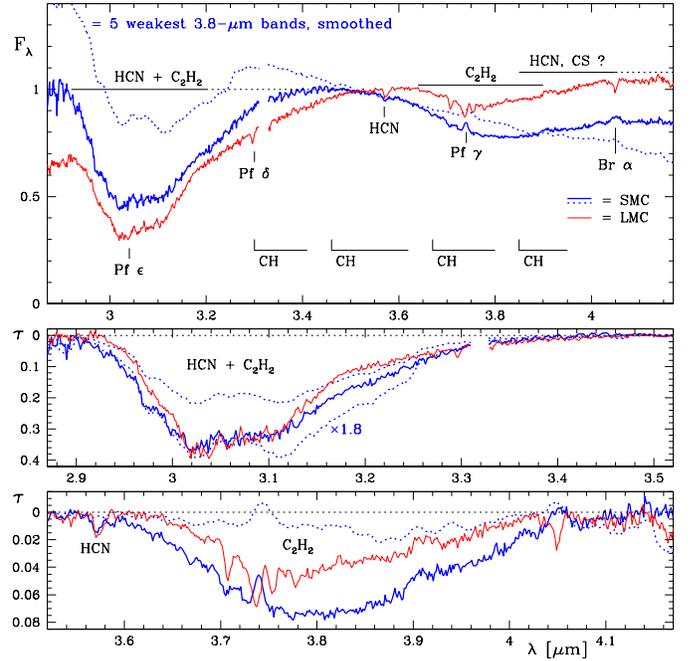,width=88mm}}
\caption[]{Average spectrum of the IR carbon stars from Table 1, excluding the
``LE'' objects in NGC\,419 which would not have been selected on their IR
properties alone (they are bluer and less dusty). A comparison is made with
the LMC (van Loon et al.\ 2006). Also shown is a smoothed average of the five
SMC stars with the weakest 3.8-$\mu$m bands (dotted). The bottom two panels
show the spectra in close-up, expressed in optical depth with respect to a
straight line connecting the spectral points at 2.87 and 3.52, and 3.52 and
4.17 $\mu$m, respectively.}
\end{figure}

%
%
\begin{table}
\caption[]{Equivalent widths (in \AA) of carbonaceous bands.}
{\scriptsize
\begin{tabular}{lrrrr}
\hline\hline
Name                 & $W_{3.1}$  & $W_{3.57}$ & $W_{3.70}$ & $W_{3.8}$  \\
\hline
{\it IR carbon stars} \\
IRAS\,00271$-$7120           & 1407     & 2.50     & 7.97     & 607      \\
GM\,780                      & 1253     & 7.73     & 1.82     & 327      \\
IRAS\,00388$-$7401           & 1358     & 0.68     & 8.87     & 499      \\
IRAS\,00393$-$7326           & 1819     & 4.94     & $-$2.03  & 971      \\
MSX\,SMC\,54                 & 1481     & 1.64     & 8.53     & 558      \\
MSX\,SMC\,44                 & 1512     & 4.91     & 3.48     & 611      \\
MSX\,SMC\,105                & 1506     & 8.95     & 3.85     & 778      \\
MSX\,SMC\,36                 & 1556     & $-$3.24  & 0.93     & 518      \\
IRAS\,00448$-$7332           & 1113     & 1.44     & 1.38     & 640      \\
MSX\,SMC\,33                 & 448      & $-$0.73  & 0.37     & 127      \\
IRAS\,00454$-$7257           & 1216     & 0.69     & 4.62     & 309      \\
                             & 1230     & 0.68     & $-$0.89  & 376      \\
IRAS\,00468$-$7341           & 1110     & 2.85     & 7.75     & 350      \\
MSX\,SMC\,66                 & 1189     & 0.64     & $-$1.68  & 234      \\
RAW\,631                     & 668      & 2.35     & 0.86     & 51       \\
MSX\,SMC\,163                & 1499     & 6.45     & 4.54     & 430      \\
ISO-MCMS\,J005149.4$-$731315 & 1325     & 0.21     & $-$0.32  & 372      \\
MSX\,SMC\,125                & 564      & 2.64     & 0.60     & $-$16    \\
GM\,106                      & 966      & $-$1.44  & 0.90     & 95       \\
MSX\,SMC\,159                & 1203     & 1.85     & 2.68     & 413      \\
LEGC\,105                    & 1435     & 1.65     & $-$1.17  & 371      \\
ISO-MCMS\,J005450.7$-$730607 & 1148     & 3.52     & 2.65     & 154      \\
ISO-MCMS\,J005454.1$-$730318 & 780      & 3.66     & 0.03     & 172      \\
MSX\,SMC\,209 & 1365         & $-$2.99  & 1.13     & 401      \\
IRAS\,00554$-$7351           & 1185     & 3.53     & 6.68     & 388      \\
ISO-MCMS\,J005720.6$-$731245 & 1073     & 0.90     & $-$0.68  & 116      \\
IRAS\,00557$-$7309           & 1328     & 0.97     & 1.52     & 457      \\
MSX\,SMC\,93                 & 1759     & $-$2.76  & 1.55     & 325      \\
NGC\,419\,LE\,16             & 1023     & $-$1.22  & 2.91     & 116      \\
NGC\,419\,IR1                & 1310     & 0.22     & $-$0.84  & 172      \\
NGC\,419\,IR2                & 1618     & $-$4.42  & 6.58     & 553      \\
NGC\,419\,LE\,35             & 1680     & 5.51     & 1.76     & 164      \\
NGC\,419\,LE\,27             & 1242     & $-$0.43  & 0.95     & 121      \\
NGC\,419\,LE\,18             & 1241     & 1.92     & 1.17     & 10       \\
GM\,522                      & 996      & 2.73     & $-$2.06  & 84       \\
IRAS\,01091$-$7320           & 1179     & 3.66     & 6.83     & 450      \\
IRAS\,01210$-$7125           & 1722     & $-$2.61  & $-$3.09  & 607      \\
\multicolumn{5}{l}{\it Oxygen-rich AGB stars and RSGs} \\
HV\,1375                     & $-$203   & 1.35     & 1.24     & 164      \\
MSX\,SMC\,18                 & $-$137   & 0.79     & 3.01     & $-$199   \\
BFM\,1                       & 55       & 0.54     & $-$0.38  & 41       \\
HV\,11262                    & $-$38    & 1.10     & 3.95     & 135      \\
PMMR\,34                     & 30       & 0.95     & 3.64     & 82       \\
IRAS\,00483$-$7347           & $-$82    & 0.85     & 2.11     & $-$261   \\
GM\,103                      & 33       & 1.47     & 1.12     & $-$230   \\
PMMR\,41                     & 31       & $-$1.72  & 0.24     & 257      \\
BMB-B\,75                    & $-$105   & 1.18     & 0.20     & $-$190   \\
HV\,1652                     & $-$549   & 3.58     & $-$0.05  & 63       \\
HV\,11417                    & 18       & 1.67     & 3.70     & $-$449   \\
IRAS\,01066$-$7332           & $-$179   & $-$3.00  & $-$0.58  & $-$517   \\
HV\,12956                    & $-$81    & $-$1.02  & 2.67     & $-$185   \\
HV\,2084                     & 8        & $-$1.33  & $-$4.34  & 423      \\
{\it R\,CrB-type stars} \\
MSX\,SMC\,14                 & 211      & $-$3.11  & $-$0.97  & $-$24    \\
MSX\,SMC\,155                & $-$54    & 0.60     & 3.94     & $-$25    \\
{\it Post-AGB objects} \\
MSX\,SMC\,29                 & $-$338   & 3.64     & 0.61     & 7        \\
IRAS\,00350$-$7436           & $-$50    & $-$0.89  & 0.95     & $-$12    \\
{\it Candidate YSOs} \\
MSX\,SMC\,79                 & 5        & 10.40    & 0.42     & $-$110   \\
IRAS\,01039$-$7305           & $-$173   & $-$1.48  & $-$0.33  & $-$35    \\
IRAS\,01042$-$7215           & 438      & 0.66     & $-$1.00  & $-$120   \\
{\it Serendipitous objects} \\
2MASS\,J00472001$-$7240350   & 144      & 0.95     & 2.55     & 112      \\
2MASS\,J00473917$-$7304115   & 1018     & 4.36     & 1.13     & 95       \\
MSX\,SMC\,79\,B              & 1546     & 18.72    & $-$24.48 & 9        \\
RAW\,616                     & $-$208   & 3.41     & $-$1.48  & $-$77    \\
RAW\,639                     & 667      & 1.40     & $-$3.68  & 90       \\
2MASS\,J00513146$-$7310513   & 45       & $-$5.54  & 7.17     & 164      \\
PMMR\,52                     & 64       & 2.01     & 2.70     & 207      \\
2MASS\,J00544039$-$7313407   & 1        & $-$4.48  & 1.37     & 46       \\
PMMR\,79 ?                   & $-$293   & 1.15     & 3.76     & 183      \\
2MASS\,J00565642$-$7335301   & 1747     & 2.20     & 12.34    & 560      \\
RAW\,1030                    & $-$288   & $-$3.05  & 6.27     & 84       \\
PMMR\,116                    & $-$244   & $-$1.72  & 3.20     & 23       \\
\hline
\end{tabular}
}
\end{table}

A mean of the spectra (after normalising to unity at 3.5 $\mu$m) was
constructed, of all carbon stars from Table 1 except the NGC\,419\,LE sources,
and compared to the equivalent spectrum for the LMC sources from van Loon et
al.\ (2006) to which we also refer the reader for an introduction to the most
prominent spectral features (Fig.\ 9). The hydrogen lines appear weakly in
emission in the SMC spectrum and in absorption in the LMC spectrum because of
the different way of correcting for the telluric standard star spectrum; these
lines should be ignored in either spectrum. There are several noteworthy
differences between the SMC and LMC spectra.

Firstly, the LMC spectrum is much redder, as a result of dust extinction and
emission. The veiling by the dust emission around 3.6--4 $\mu$m may be the
cause for a weaker 3.8-$\mu$m band in the LMC compared to the SMC, whilst the
3.1-$\mu$m band appears approximately equally strong (see also Ohnaka et al.\
2007).

%
%
\begin{figure*}
\centerline{\vbox{
\hbox{
\psfig{figure=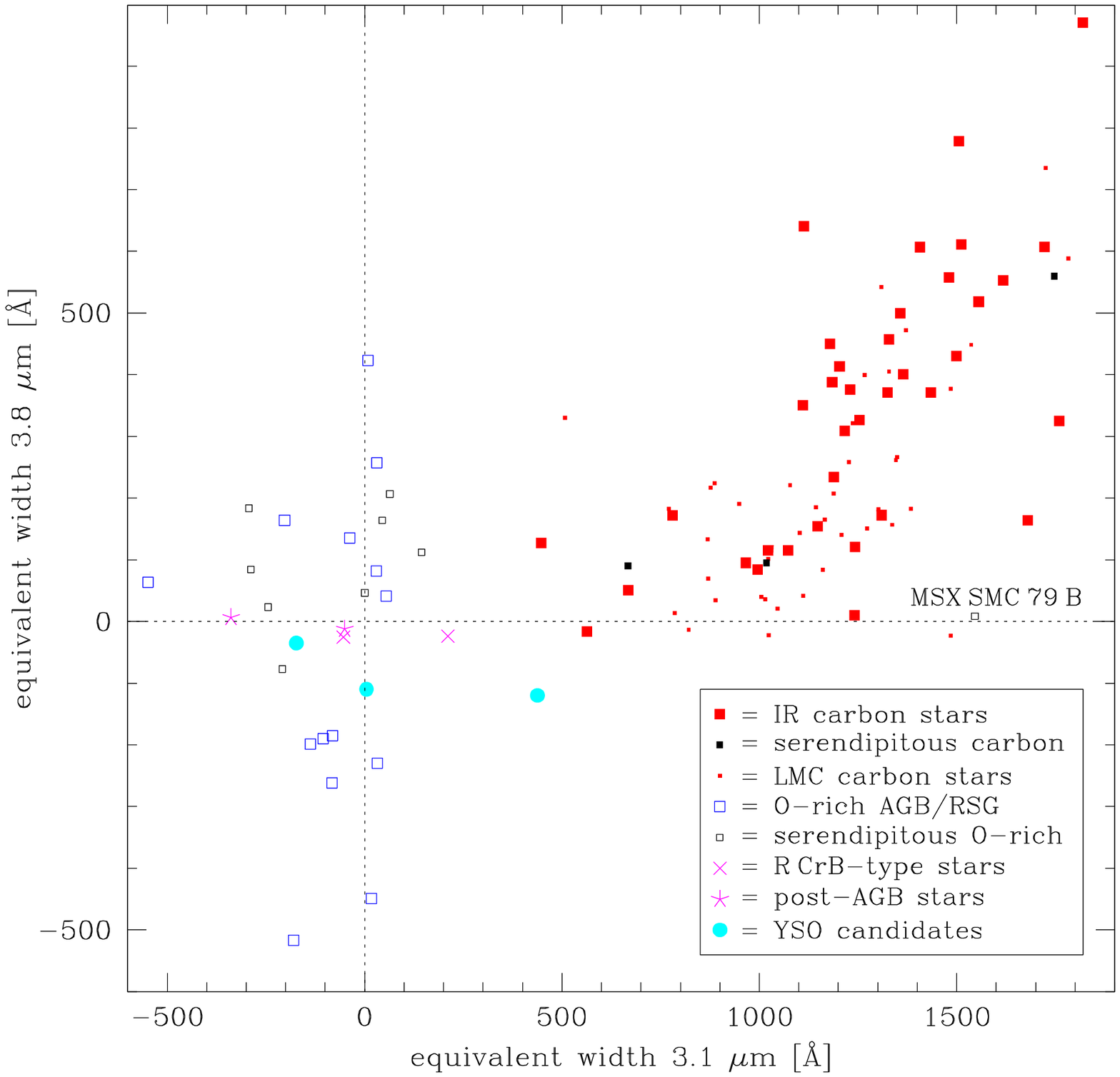,width=88mm}\hspace{4mm}
\psfig{figure=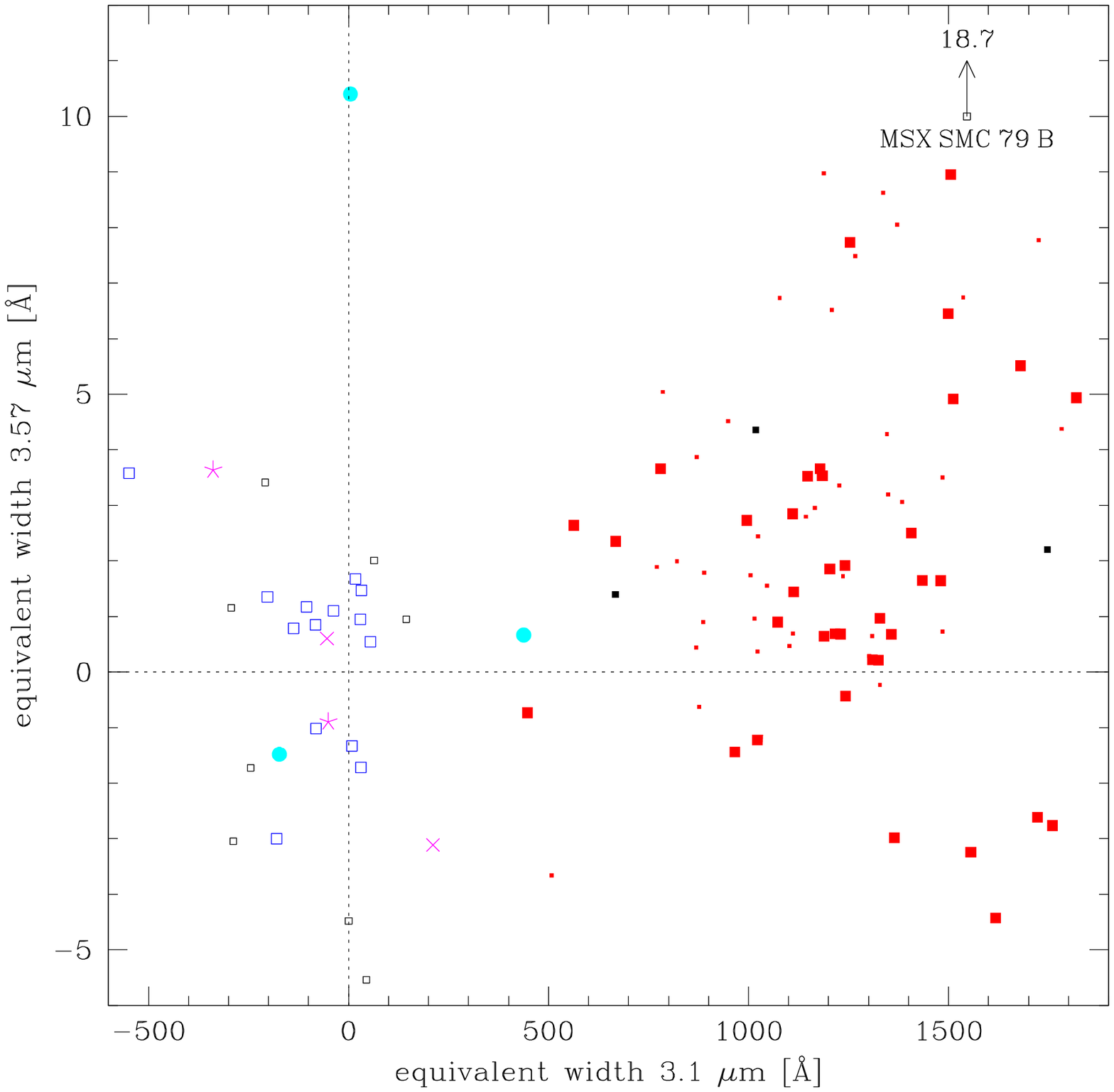,width=88mm}}
\vspace{2mm}
\hbox{
\psfig{figure=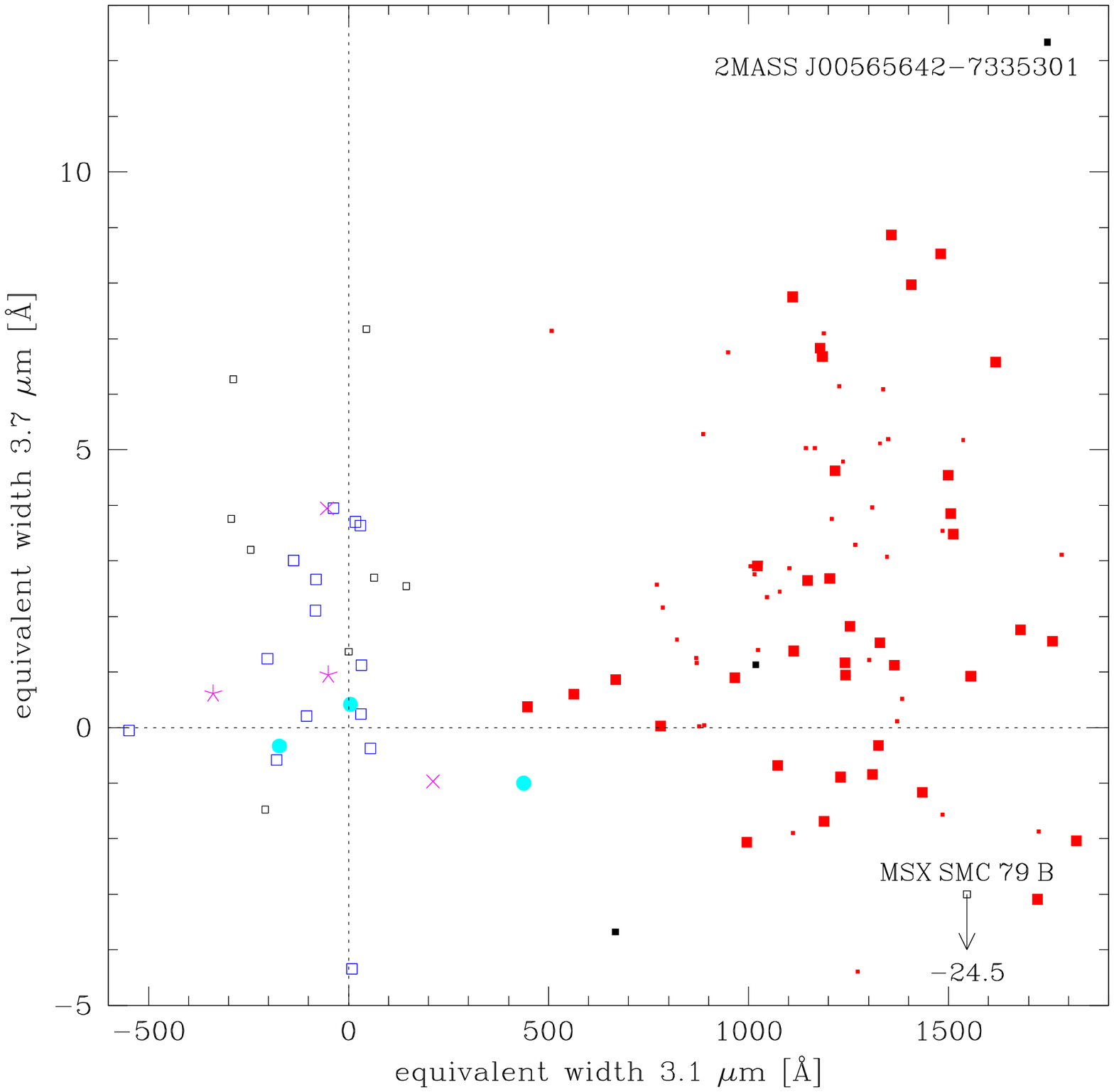,width=88mm}\hspace{4mm}
\psfig{figure=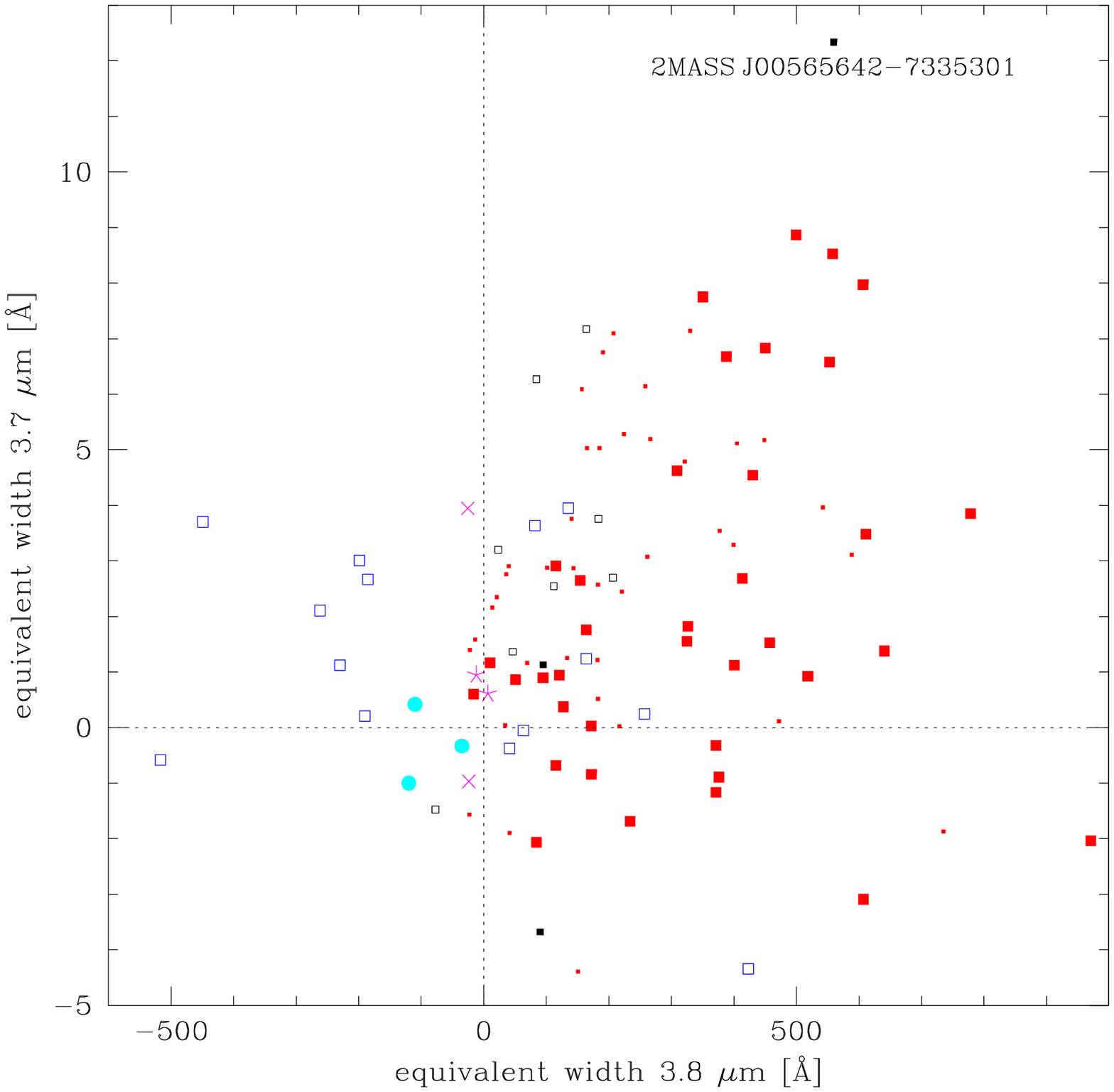,width=88mm}}
}}
\caption[]{Equivalent widths (in \AA) of the 3.1-, 3.57-, 3.70- and 3.8-$\mu$m
molecular bands. Carbon stars in the SMC and LMC show identical trends and
cover identical ranges, apart from a slight difference in the W$_{3.7}$ versus
$W_{3.8}$ diagram. The other objects mostly scatter around zero, as expected.}
\end{figure*}

Secondly, the 3.8-$\mu$m acetylene (C$_2$H$_2$) band in the SMC spectrum peaks
slightly longward of 3.80 $\mu$m whereas in the LMC spectrum it peaks slightly
shortward of 3.80 $\mu$m. This may be related to the difference in the
3.70-$\mu$m feature, which is very weak in the SMC compared to the LMC. This
is one of a few sharp features attributed to acetylene at low pressure (colder
gas at higher elevation --- van Loon et al.\ 2006), and may thus help unravel
the connection between the dust formation and the underlying molecular
atmosphere. In the LMC, the 3.70-$\mu$m feature correlates with a ``knee''
between 3.15--3.20 $\mu$m (van Loon et al.\ 2006), which is indeed also weaker
in the SMC.

Thirdly, the HCN band at 3.57 $\mu$m is weaker in the SMC than in the LMC,
possibly as a result of a lower nitrogen abundance in the metal-poor SMC. On
the other hand, it appears as if the satellite band of HCN at 3.58--3.59
$\mu$m is relatively strong in the SMC carbon stars.

The average of the five SMC carbon stars with the weakest 3.8-$\mu$m bands is
shown for comparison (Fig.\ 9, dotted). These stars have an even bluer
continuum than the SMC sample as a whole, suggesting that dust is largely
absent in the absence of the 3.8-$\mu$m band. Note that some weak absorption
is still discernible near 3.9 $\mu$m. The 3.1-$\mu$m band is also weaker, but
the shape is similar to that of the other SMC stars; when scaled by a factor
1.8 (Fig.\ 9, middle panel) the profiles match up very well except that the
stars with weak 3.8-$\mu$m absorption have a deeper long-wavelength side to
the 3.1-$\mu$m band, possibly as the molecular atmosphere is less extended and
warmer.

\subsubsection{Molecular band strengths}

The equivalent widths of the 3.1-, 3.57-, 3.70- and 3.8-$\mu$m bands were
measured following their definitions by van Loon et al.\ (2006) (Table 5).
This was done for all the objects in Tables 1--4, not just the carbon stars.
The relations between the various band strengths are explored in Fig.\ 10.
Clearly, the carbon stars all have strong 3.1-$\mu$m bands which separates
them from the non-carbon stars. The only interloper is the noisy spectrum of
the YSO candidate MSX\,SMC\,79\,B. The scatter of the non-carbon stars around
zero gives an indication of the typical errorbars in the individual equivalent
width measurements.

The 3.8-$\mu$m band becomes visible in carbon stars with $W_{3.1}\gsim1000$
\AA, and then correlates with 3.1-$\mu$m band strength (Fig.\ 10). Remarkably,
this trend is indistinguishable between the SMC and LMC. Equally remarkable,
the SMC and LMC objects span the same range in equivalent width, for all bands
concerned. This is even true for the 3.57- and 3.70-$\mu$m bands, although
there are fewer SMC carbon stars with substantial ($W\gsim4$ \AA) HCN
absorption at 3.57 $\mu$m and acetylene absorption at 3.70 $\mu$m than in the
LMC, resulting in weaker features in the average SMC spectrum of Fig.\ 9; some
have negative values, resulting from inaccuracies in defining the continuum in
noisy spectra with weak (or absent) absorption features.

The stars with the strongest 3.57-$\mu$m HCN band are also the ones with the
strongest 3.1-$\mu$m band, which has a contribution from HCN (Fig.\ 10); the
same is seen in the LMC. However, the reverse is not true: the most negative
values for the 3.57-$\mu$m band are amongst the strongest 3.1-$\mu$m absorbers
too (Fig.\ 10); this is not seen in the LMC. Higher resolution and lower noise
levels are required before interpreting the negative values as the HCN band
appearing in emission.

The 3.70-$\mu$m acetylene band is stronger in stars with strong 3.1-$\mu$m
bands, which has a contribution from acetylene (Fig.\ 10). However, the stars
with the strongest 3.1-$\mu$m bands do not show the sharp 3.70-$\mu$m feature.
An identical trend is seen in the SMC and LMC.

The SMC and LMC stars differ in the relation between the sharp 3.70- and broad
3.8-$\mu$m acetylene bands (Fig.\ 10): the 3.70-$\mu$m band strength peaks at
a larger 3.8-$\mu$m band strength in the SMC. Veiling by dust emission would
affect both bands about equally, so there must be a more delicate difference
between the warmer lower layers and the colder elevated layers in the
molecular atmosphere. The SMC stars seem to have more trouble in getting the
acetylene to higher elevations than the LMC stars. Alternatively, the
3.70-$\mu$m band may form throughout the dust envelope, in which case the
difference is merely the result of stronger veiling of the 3.8-$\mu$m band in
the dustier LMC stars compared to the less dusty SMC stars. IRAS\,00393$-$7326
has the strongest 3.1- and 3.8-$\mu$m bands and is not particularly red
($K_{\rm s}-L^\prime_{\rm NB}=2.6$ mag), but it does not show any 3.70-$\mu$m
absorption. It does show a clear 3.57-$\mu$m HCN absorption feature.

%
%
\begin{figure}
\centerline{\psfig{figure=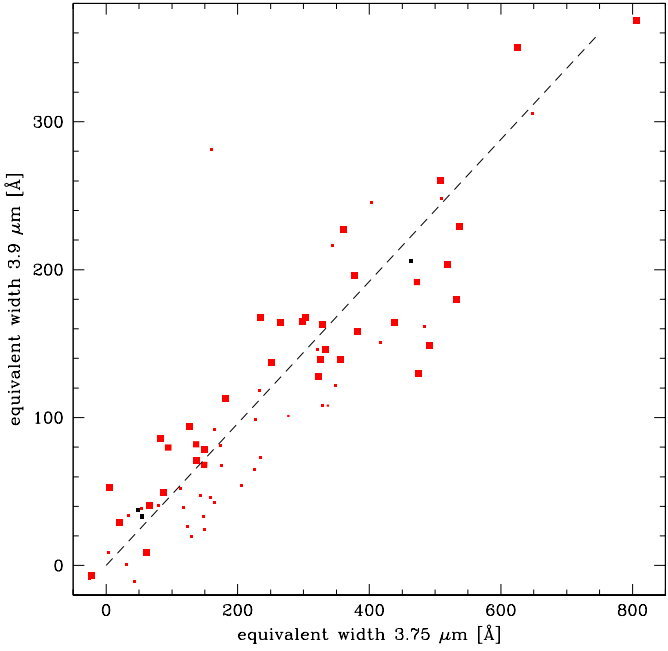,width=88mm}}
\caption[]{Equivalent widths (in \AA) of the 3.75- and 3.9-$\mu$m molecular
bands (symbols are as in Fig.\ 10). The dashed line is for reference only.}
\end{figure}

The 3.8-$\mu$m band may have a contribution from other molecules than
acetylene, in particular molecules with C--H bonds and possibly carbon sulfide
(CS) around 3.9 $\mu$m. To investigate this, the 3.8-$\mu$m band is split up
into a 3.75-$\mu$m band running from 3.60 to 3.90 $\mu$m, and a 3.9-$\mu$m
band running from 3.85 to 4.03 $\mu$m (note the slight overlap). The continuum
is determined in both cases by a straight line in the F$_\lambda$ spectrum
between the medium flux density at 3.50--3.60 $\mu$m and that at 4.03--4.10
$\mu$m. We find that in fact both bands correlate extremely well (Fig.\ 11),
suggesting that the contribution from other molecules, in particular CS, may
be small. The SMC and LMC stars show the same behaviour, apart from a slight
offset at weak bandstrengths --- this could be due to a contribution to the
3.75-$\mu$m band from the 3.70-$\mu$m absorption components, which peak at
lower 3.8-$\mu$m bandstrength in the LMC than in the SMC as discussed above.
The LMC outlier is BMB-R\,46 (SHV\,0521050$-$690415; Matsuura et al.\ 2005),
whose spectrum is extremely noisy around 4 $\mu$m.

\subsubsection{Molecules and dust}

%
%
\begin{figure}
\centerline{\psfig{figure=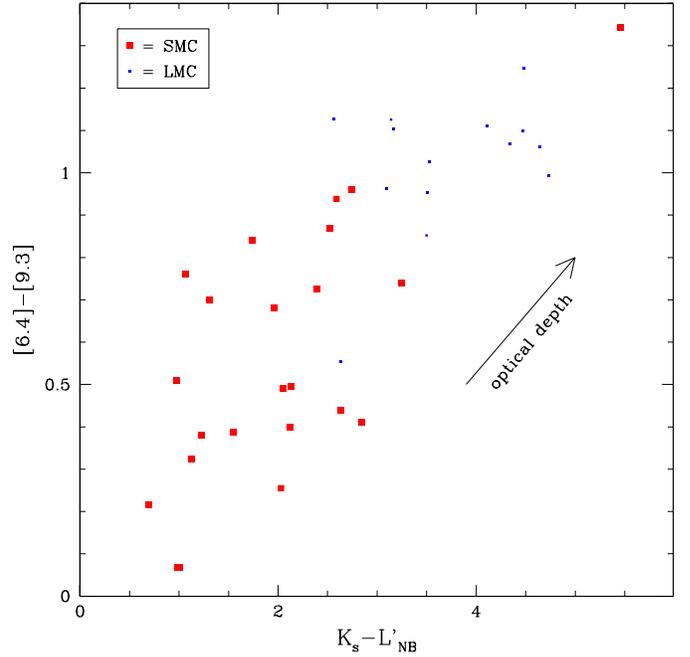,width=88mm}}
\caption[]{Correlation for carbon stars, between the K$_{\rm
s}$--L$^\prime_{\rm NB}$ colour and the [6.4]--[9.3] colour defined for
Spitzer IRS spectra by Sloan et al.\ (2006) and Zijlstra et al.\ (2006). The
latter is an accurate measure of the optical depth.}
\end{figure}

The dust envelope causes extinction below 4 $\mu$m and possibly emission from
3 $\mu$m onwards. The K--L colour can be used as a proxy for the optical depth
of the dust envelope: Fig.\ 12 shows a good correlation between the K$_{\rm
s}$--L$^\prime_{\rm NB}$ colour and the [6.4]--[9.3] colour defined in Sloan
et al.\ (2006) and Zijlstra et al.\ (2006) for the carbon stars with published
Spitzer IRS spectra. These authors show that the [6.4]--[9.3] colour is an
accurate measure of the dust optical depth. By inference, the K--L colour is
hereby proven to be a good measure of the dust optical depth too. Although a
few SMC carbon stars are very red, the LMC carbon stars are generally
significantly redder, despite similar molecular band strength (Fig.\ 13). This
suggests that although SMC and LMC stars are similarly proficient in producing
molecules, the metal-poor SMC stars are less able to form dust from it.

The dustiest LMC stars clearly have relatively weak 3.8-$\mu$m bands (Fig.\
13), which is evidence of the veiling effect that the dust continuum emission
has on the equivalent width of the absorption band. This shows that most of
the 3.8-$\mu$m band is formed in the warmer layers below the dust envelope.
Emission from the molecular bands themselves in strongly pulsating stars can
add to the veiling (Hron et al.\ 1998; Ohnaka et al.\ 2007). The same may
explain the 3.1-$\mu$m band, which is only moderately strong in the reddest
carbon stars as opposed to the large spread in band strengths observed amongst
bluer carbon stars (Fig.\ 13) --- the latter is also seen in the Milky Way
(e.g., Le Bertre et al.\ 2005).

On the other hand, the 3.70-$\mu$m band clearly reaches maximum strength in
the dustier objects, both in the SMC and LMC, which suggests an intimate
relation between the dust and the colder acetylene gas at high elevation. This
reinforces our previous observation (Section 4.1) that dust is not apparent in
stars that show little acetylene, and suggests acetylene as a prime building
block for carbonaceous grain formation. The star with the strongest
3.70-$\mu$m band is 2MASS\,J00565642$-$7335301; it is not red at all, but its
(serendipitous) spectrum is rather noisy and the 3.70-$\mu$m band may have
been overestimated. Veiling might also be the reason why the stars with the
strongest 3.57-$\mu$m band are the least dusty (Fig.\ 13). This suggests that
the HCN responsible for the 3.57-$\mu$m feature is present predominantly close
to the stellar surface.

%
%
\begin{figure*}
\centerline{\vbox{
\hbox{
\psfig{figure=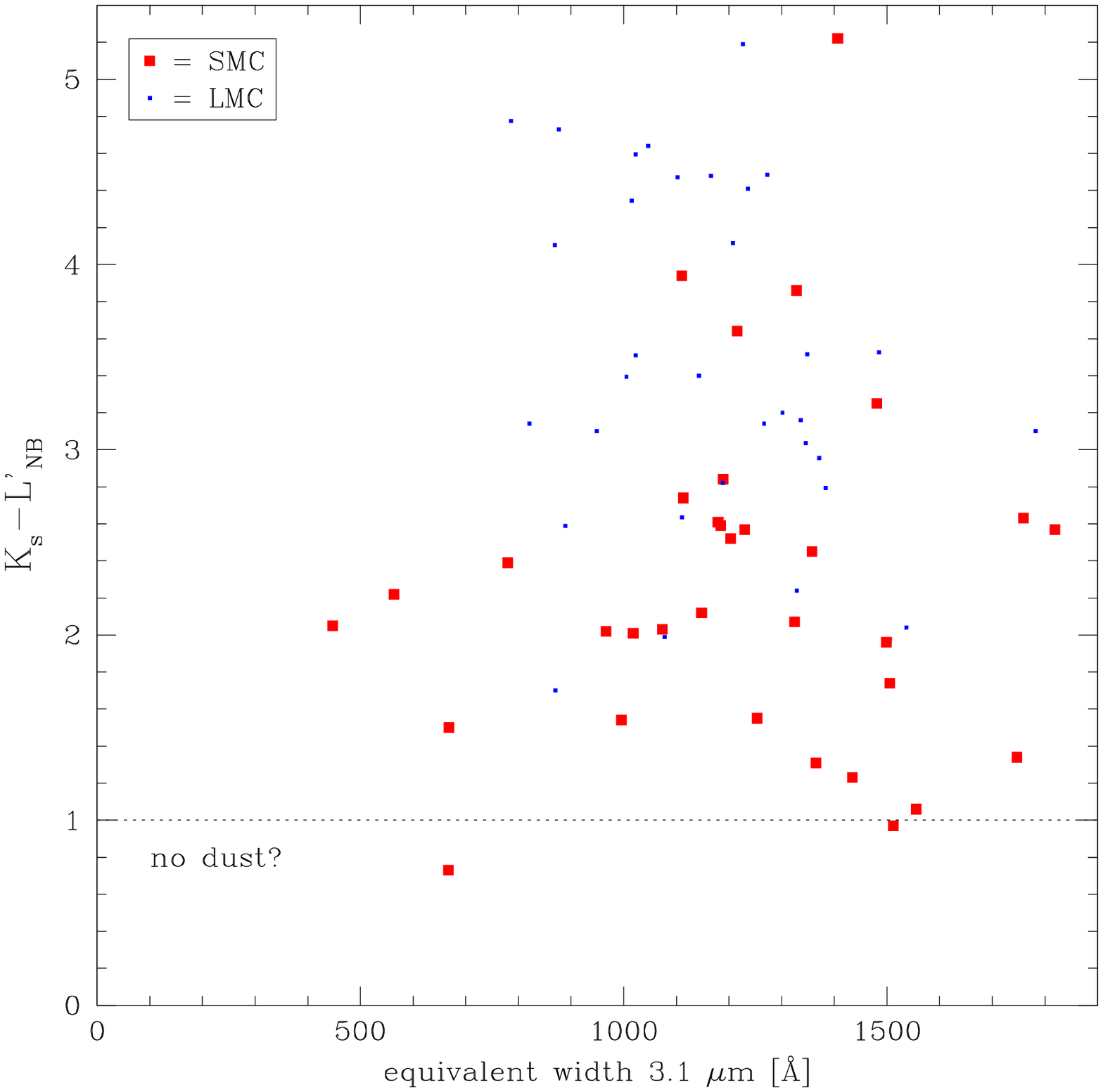,width=88mm}\hspace{4mm}
\psfig{figure=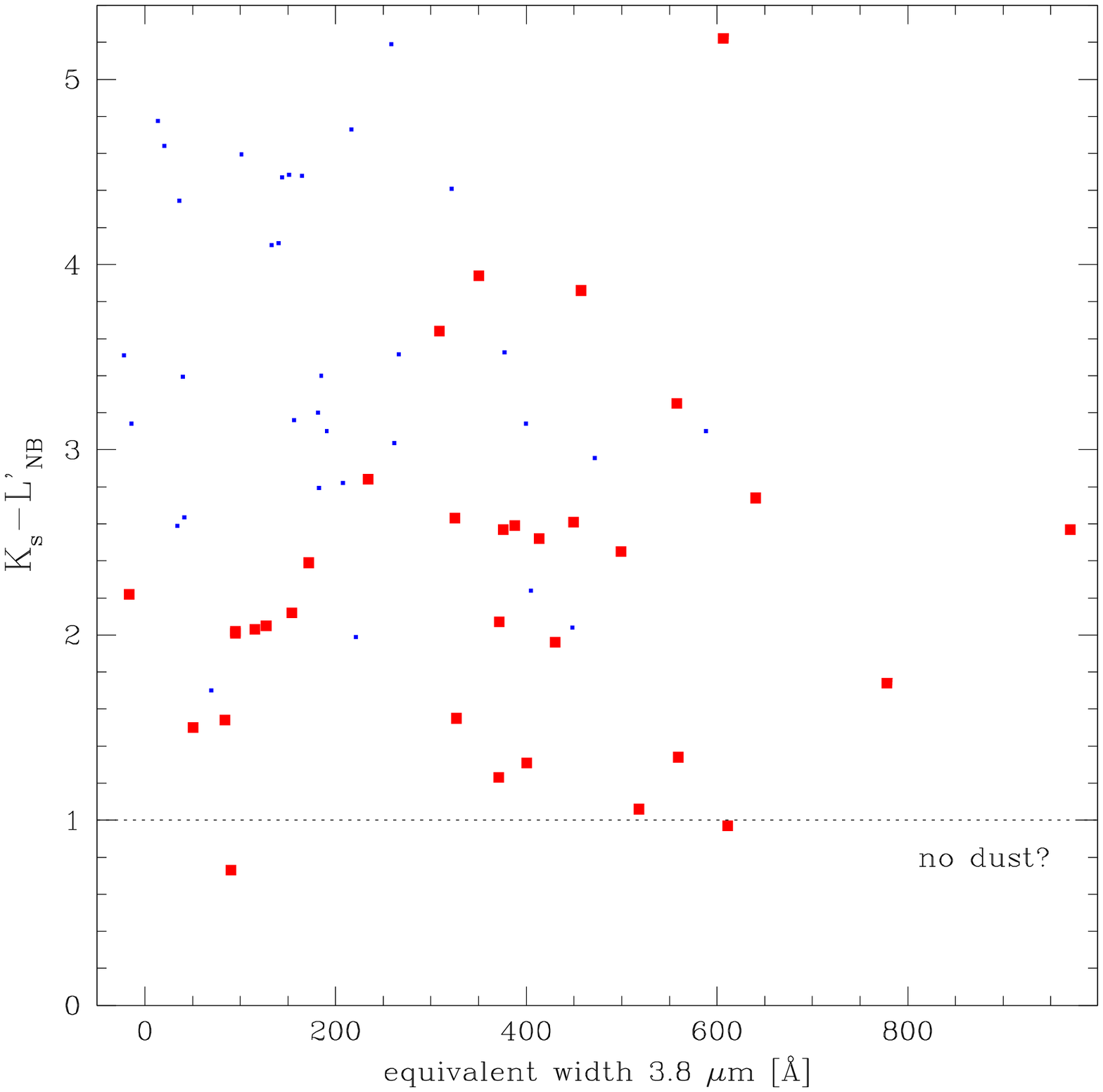,width=88mm}}
\vspace{2mm}
\hbox{
\psfig{figure=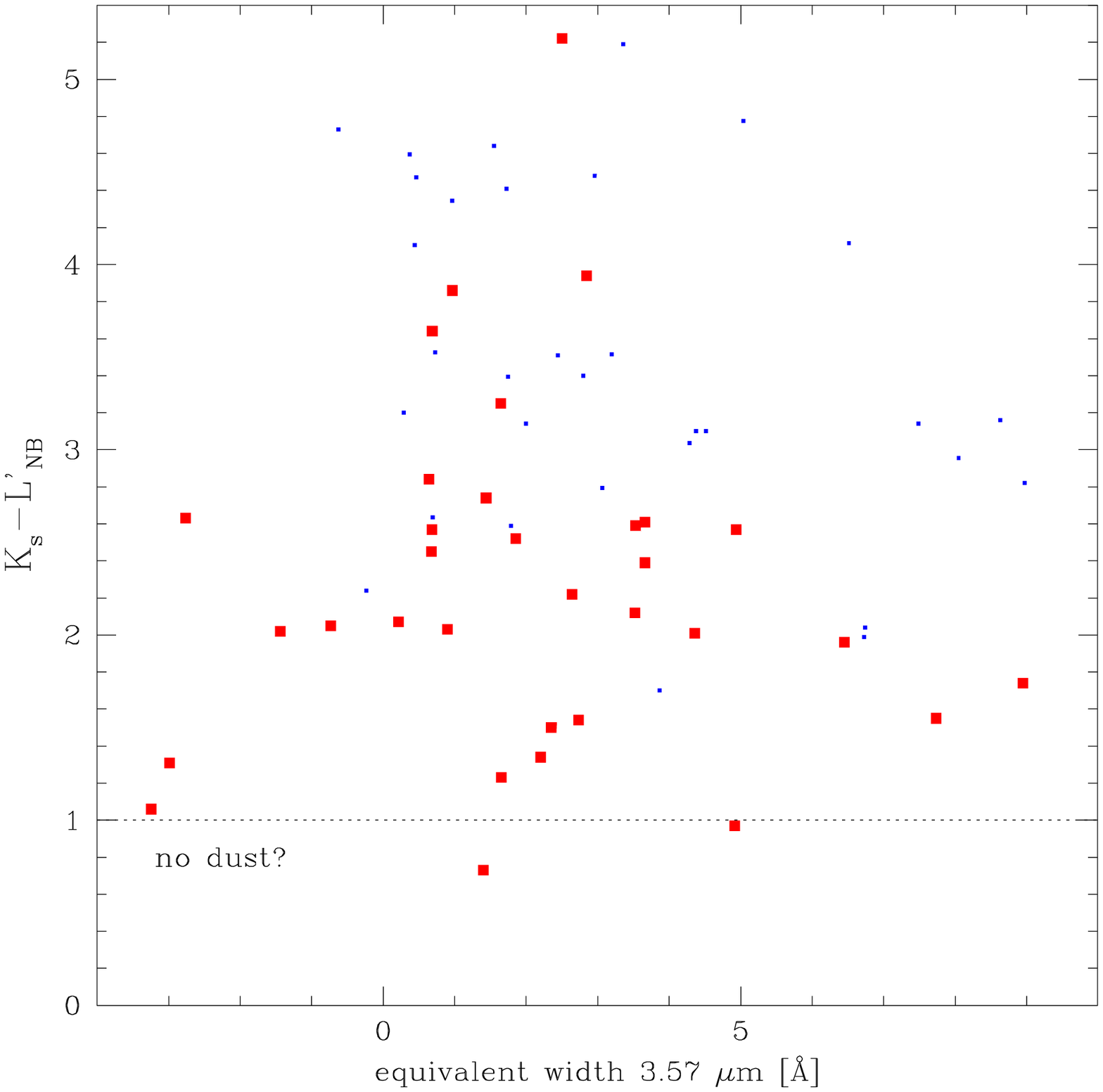,width=88mm}\hspace{4mm}
\psfig{figure=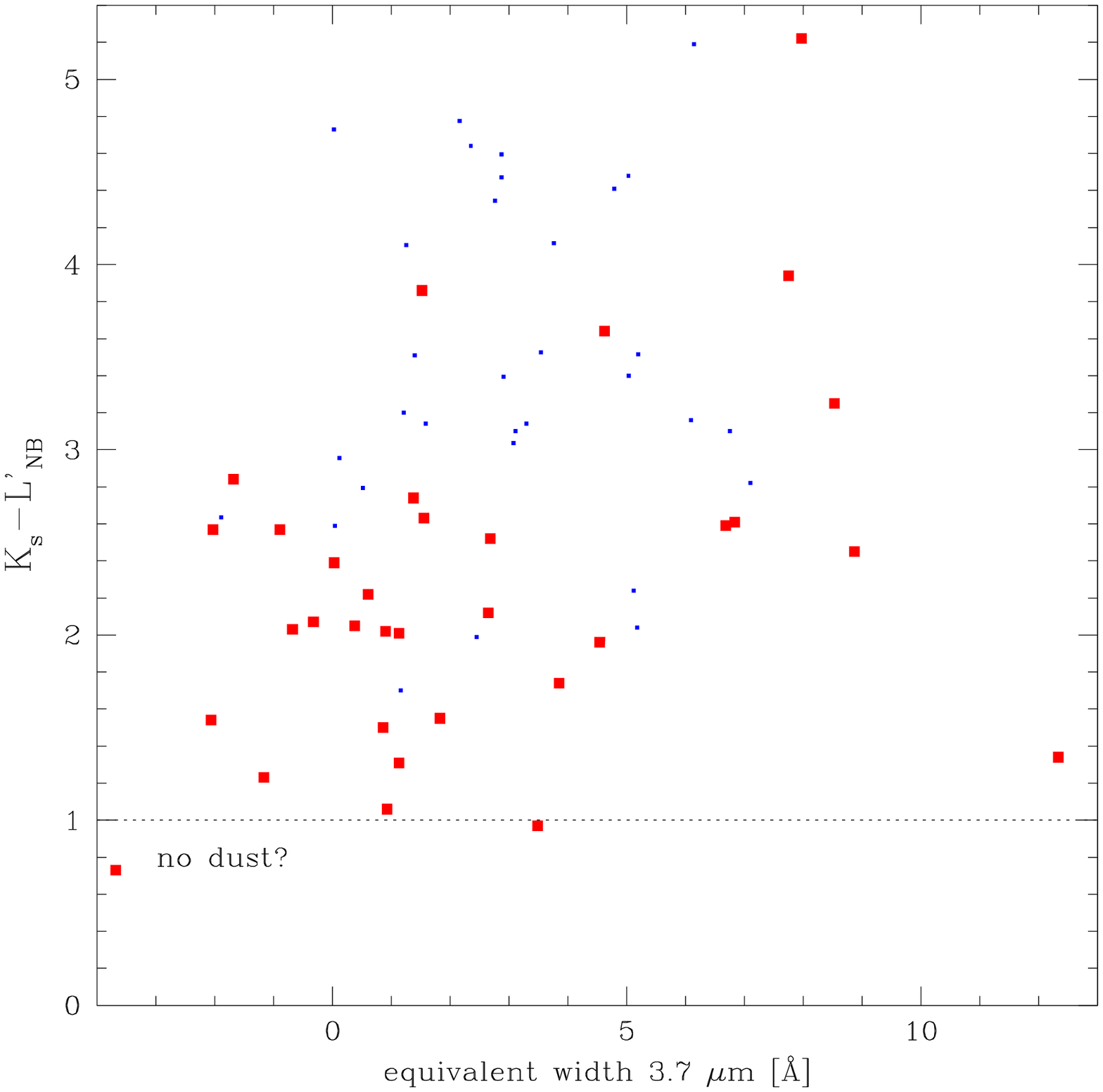,width=88mm}}
}}
\caption[]{Dependence of the K$_{\rm s}$--L$^\prime_{\rm NB}$ colour, a proxy
for the optical depth of the circumstellar dust envelope, and the equivalent
widths (in \AA) of the 3.1-, 3.57-, 3.70- and 3.8-$\mu$m molecular bands. The
differences between carbon stars in the SMC and LMC may reflect a different
connection between the gas and the dust. Stars from the NGC\,419 cluster are
not included (see Fig.\ 13).}
\end{figure*}

\subsubsection{Evolutionary sequence in NGC\,419}

NGC\,419 offers known initial masses and metallicity. We can place the cluster
members in an evolutionary sequence, as on the thermal-pulsing AGB differences
in time are more important than differences in initial mass. Assuming that the
thickest dust envelopes are formed in the final stages of AGB evolution, the
K--L colour can serve as a proxy for evolutionary stage (``time'').

Despite the small number of objects, the trends in the band strengths with
time confirm our previous analysis (Fig.\ 14). The 3.57-$\mu$m band is only
discernible in one of the least dusty stars, and then disappears. The
3.1-$\mu$m band shows little correlation; this may be due to the fact that the
band is saturated and both HCN and acetylene contribute. On the other hand,
the 3.8-$\mu$m band is strong only in the dustiest star. The 3.70-$\mu$m band
also becomes stronger as the dust envelope thickens. This suggests that, as
the molecular atmosphere thickens (as a result of lower temperature and/or
higher C/O ratio), more acetylene is brought to higher elevation and more dust
is formed. The elevation of the molecular atmosphere is probably caused by the
stronger pulsation as the star evolves (and cools).

The two stars, IR1 and IR2, in the similar cluster NGC\,1978 in the LMC (van
Loon et al.\ 2005b, 2006) (small symbols in Fig.\ 14) confirm exactly the same
behaviour as seen in NGC\,419. This is not surprising, as de Freitas Pacheco
et al.\ (1998) measured identical metallicities of [Fe/H]$=-0.60\pm0.21$ for
both clusters, although they find that NGC\,419, at $1.2\pm0.5$ Gyr, is
somewhat younger than NGC\,1978, at $3.0\pm1.0$ Gyr. The associated difference
in mass of the AGB stars in these clusters apparently is not large enough to
lead to noticeable differences in their molecular atmospheres.

\subsubsection{Polycyclic Aromatic Hydrocarbons}

The post-AGB object MSX\,SMC\,29 shows the best example of the 3.28-$\mu$m
band attributed to emission from PAHs (Fig.\ 5), stronger than in the other
SMC post-AGB object, IRAS\,00350$-$7436 (Matsuura et al.\ 2005). Its K$_{\rm
s}$--L$^\prime_{\rm NB}$ colour is somewhat bluer than the loosely defined
sequence of dust-enshrouded carbon stars (Fig.\ 1); it is also bluer than the
two R\,CrB-type stars in our sample, and brighter in the near-IR than the
reddest of these two. The central star of IRAS\,00350$-$7436 has a G- or
K-type optical spectrum with strong CN bands, and it is about as luminous as
AGB stars get, $L\sim50,000$ K (Whitelock et al.\ 1989; Groenewegen et al.\
2000) --- the faint near-IR magnitudes are a result of the relatively large
near-IR bolometric corrections for this warm star. This suggests a relatively
massive AGB star progenitor, which is intriguing given that it is a
carbon-rich object. HBB prevents massive AGB stars, with initial masses in
excess of $\sim4$ M$_\odot$, from becoming a carbon star (Boothroyd, Sackmann
\& Ahern 1993). On the other hand, once mass loss has almost completely
depleted the stellar mantle, HBB will cease and another thermal pulse could
still turn the star into a carbon star (Frost et al.\ 1998). The latter
scenario predicts that the carbon-rich dust shell is enveloped by a much
older, cooler oxygen-rich dust shell.

%
%
\begin{figure}
\centerline{\psfig{figure=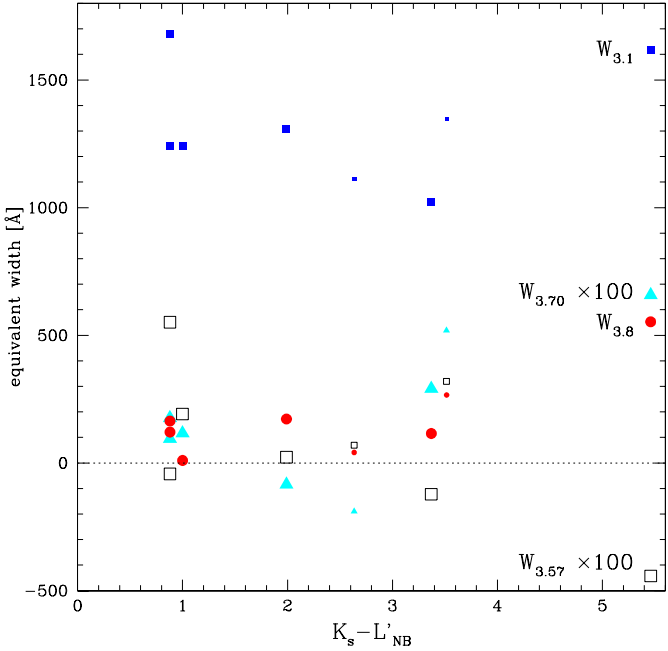,width=88mm}}
\caption[]{Equivalent widths (in \AA) of the 3.1-, 3.57-, 3.70- and 3.8-$\mu$m
molecular bands as a function of the K$_{\rm s}$--L$^\prime_{\rm NB}$ colour,
in the populous intermediate-age cluster NGC\,419, and (small symbols) in the
similar cluster NGC\,1978 in the LMC (van Loon et al.\ 2005b, 2006). The
thickest circumstellar dust shells show stronger acetylene absorption, both at
3.8- and 3.70 $\mu$m.}
\end{figure}

MSX\,SMC\,29 was observed with the Spitzer IRS by Kraemer et al.\ (2006), who
detect several PAH emission bands in the 6--13 $\mu$m region, as well as
absorption from acetylene and larger hydrocarbons. They interpret the presence
of aliphatic bonds as evidence of a mild ionization environment, placing the
object to within only a few centuries since its departure from the AGB.
Aliphatic bonds are expected to produce additional PAH emission at 3.4 $\mu$m
(Sloan et al.\ 2007); this is not seen in our L-band spectrum (Fig.\ 5), but
the spectrum does show a depression where this feature is expected, suggesting
that it might be in absorption. Such an unusual result requires confirmation
with higher signal-to-noise data.

R\,CrB-type stars are hydrogen-poor, and believed to sporadically form
carbon-rich dust (Clayton 1996), whilst the heating central stars of post-AGB
objects have stopped forming fresh dust. R\,CrB-type stars may well be the
``proto-post-AGB'' objects that are just leaving the AGB. There may thus exist
an evolutionary sequence from optically bright carbon stars $\rightarrow$
dust-enshrouded carbon stars $\rightarrow$ R\,CrB-type stars $\rightarrow$
carbon-rich post-AGB objects, where the PAHs appear as the central star of the
post-AGB object becomes hot enough to destroy the dust grains that were formed
not long before during the AGB and/or R\,CrB phase, and to excite the
energetic transitions of the PAHs (cf.\ Sloan et al.\ 2007). However, recent
measurements of greatly enhanced $^{18}$O in R\,CrB stars have been taken as
evidence against a very late helium-flash scenario (such as occurring on the
white dwarf cooling track), possibly pointing instead at a double-degenerate
merger origin (Clayton et al.\ 2007).

\subsection{Oxygen chemistry}

The cooler, M-type stars show a curvature resulting from molecular absorption,
mainly of water (H$_2$O) and silicon monoxide (SiO) (e.g., Tsuji et al.\ 1997;
Yamamura, de Jong \& Cami 1999; Decin \& Eriksson 2007). A simple way of
quantifying this is obtained by averaging the flux densities at 3 and 4
$\mu$m, and comparing it to the flux density at 3.5 $\mu$m:
\begin{equation}
S_{\rm mol} = 2 F_{3.5} / (F_3+F_4) - 1
\end{equation}
A strongly curved spectrum due to abundant molecules yields a positive value
for this ``molecular index'', whilst a spectrum with little molecular
absorption yields a negative value. The computed values are listed in Table 6.

%
%
\begin{table}
\caption[]{Oxygenous band strengths in AGB stars and RSGs. The SiO band may
appear in absorption (A) or emission (E).}
{\scriptsize
\begin{tabular}{lrrll}
\hline\hline
Name  & $S_{\rm mol}$ & $W_{\rm OH}$ & SiO & SpT \\
\hline
{\it SMC} \\
HV\,1375                   &    0.03 &    1.91 & E+A & M5      \\
MSX\,SMC\,18               &    0.12 &    0.44 & A?  &         \\
BFM\,1                     & $-$0.06 &    0.36 &     & S       \\
HV\,11262                  & $-$0.04 &    3.66 &     & M0.5    \\
PMMR\,34                   & $-$0.10 &    4.32 & E?  & M0.5    \\
IRAS\,00483$-$7347         &    0.22 &    2.77 & A?  & M8      \\
GM\,103                    &    0.21 &    0.89 & A?  & M4      \\
PMMR\,41                   & $-$0.15 &    3.67 & E   & M0\,Iab \\
BMB-B\,75                  &    0.08 &    0.32 & A   & M       \\
HV\,1652                   & $-$0.09 &    1.40 & E   & M0.5    \\
HV\,11417                  &    0.16 &    2.08 & E?  & M5\,e   \\
IRAS\,01066$-$7332         &    0.30 & $-$0.90 &     & M8      \\
HV\,12956                  &    0.15 &    1.41 &     & M5      \\
HV\,2084                   & $-$0.10 & $-$0.53 & E   & M2\,Ia  \\
2MASS\,J00472001$-$7240350 & $-$0.13 &    2.69 & ?   &         \\
2MASS\,J00513146$-$7310513 & $-$0.14 &    0.65 & A   &         \\
PMMR\,52                   & $-$0.14 &    4.09 & E   & K5      \\
2MASS\,J00544039$-$7313407 & $-$0.05 &    5.37 &     &         \\
PMMR\,79 ?                 &    0.14 &    3.30 & E?  & K5      \\
PMMR\,116                  & $-$0.02 &    1.84 & A   & K6.5    \\
{\it LMC} \\
IRAS\,05280$-$6910         & $-$0.04 & $-$2.10 & E+A & OH/IR   \\
NGC\,1984\,IR2             & $-$0.06 &    3.70 & ?   &         \\
MSX\,LMC\,50\,B            & $-$0.09 &    3.66 & A   &         \\
HV\,11977                  & $-$0.10 &    2.86 & A   & M       \\
HV\,2532                   &    0.00 &    4.00 & A   & M4\,I   \\
MSX\,LMC\,1780\,B          & $-$0.13 &    4.46 &     &         \\
IRAS\,05128$-$6455         & $-$0.16 & $-$0.29 &     & M9      \\
IRAS\,05148$-$6730         & $-$0.19 &    1.20 &     & M2.5\,I \\
\hline
\end{tabular}
}
\end{table}

%
%
\begin{figure*}
\centerline{\hbox{
\psfig{figure=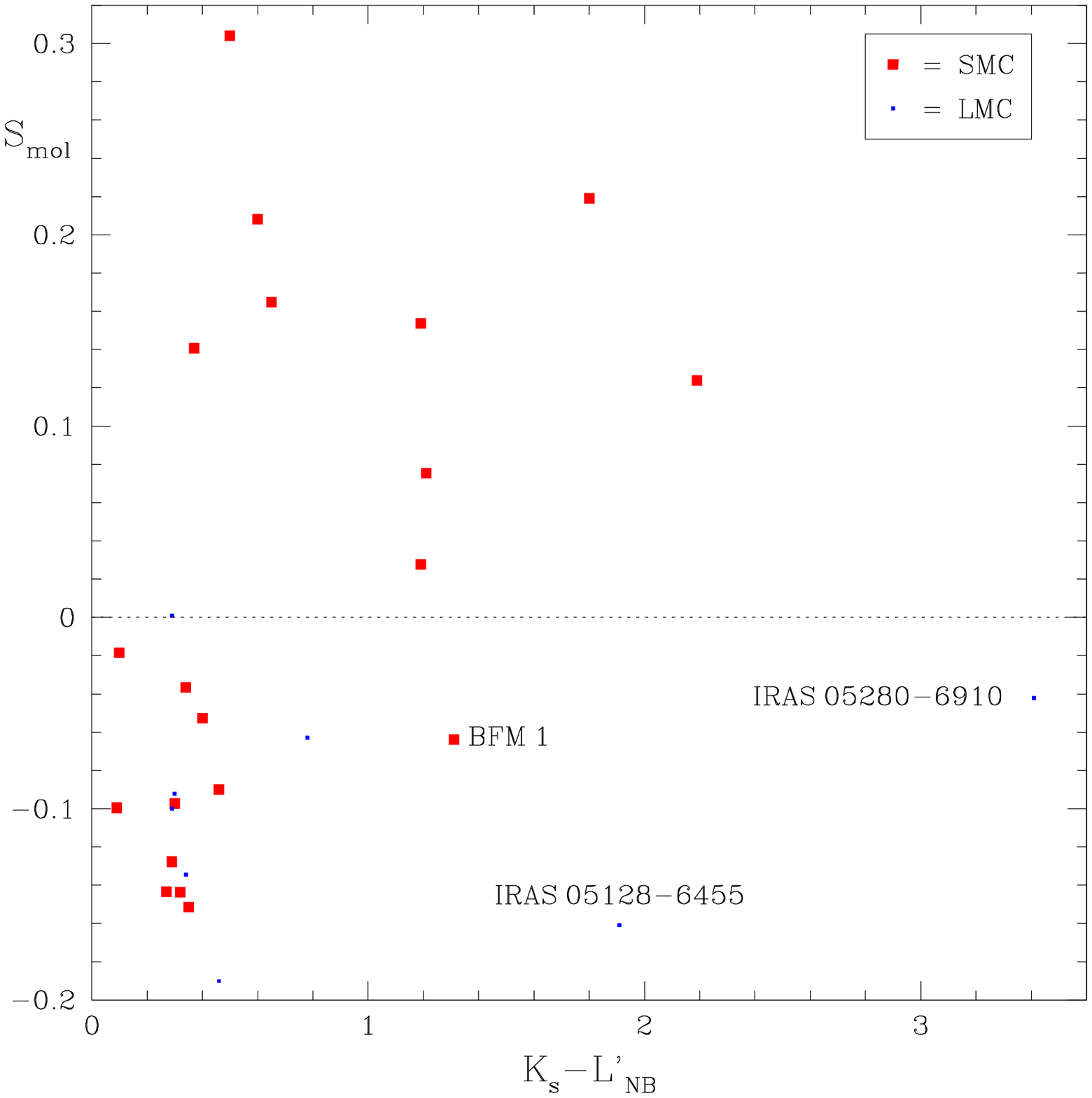,width=58mm}\hspace{2mm}
\psfig{figure=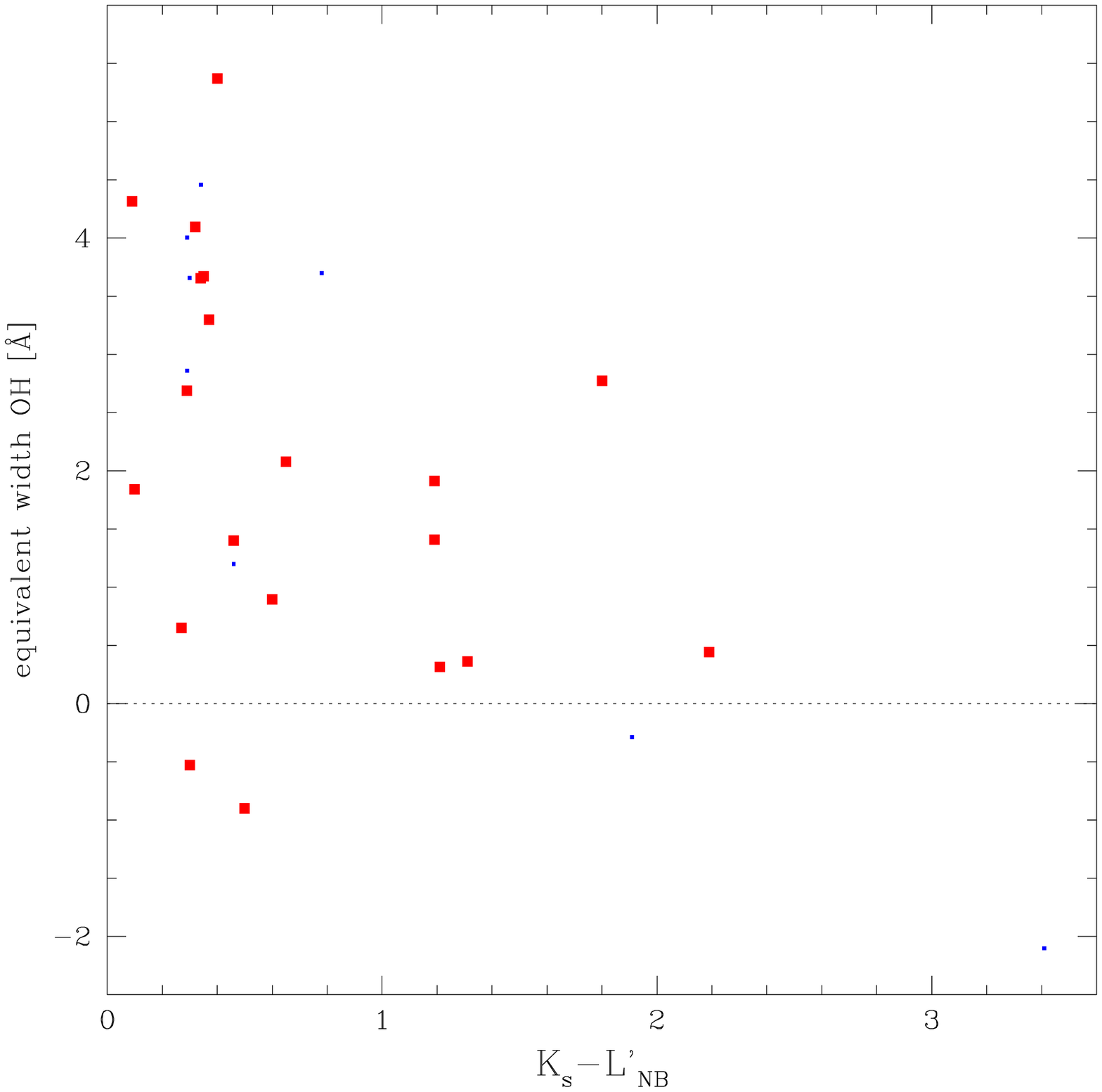,width=59mm}\hspace{2mm}
\psfig{figure=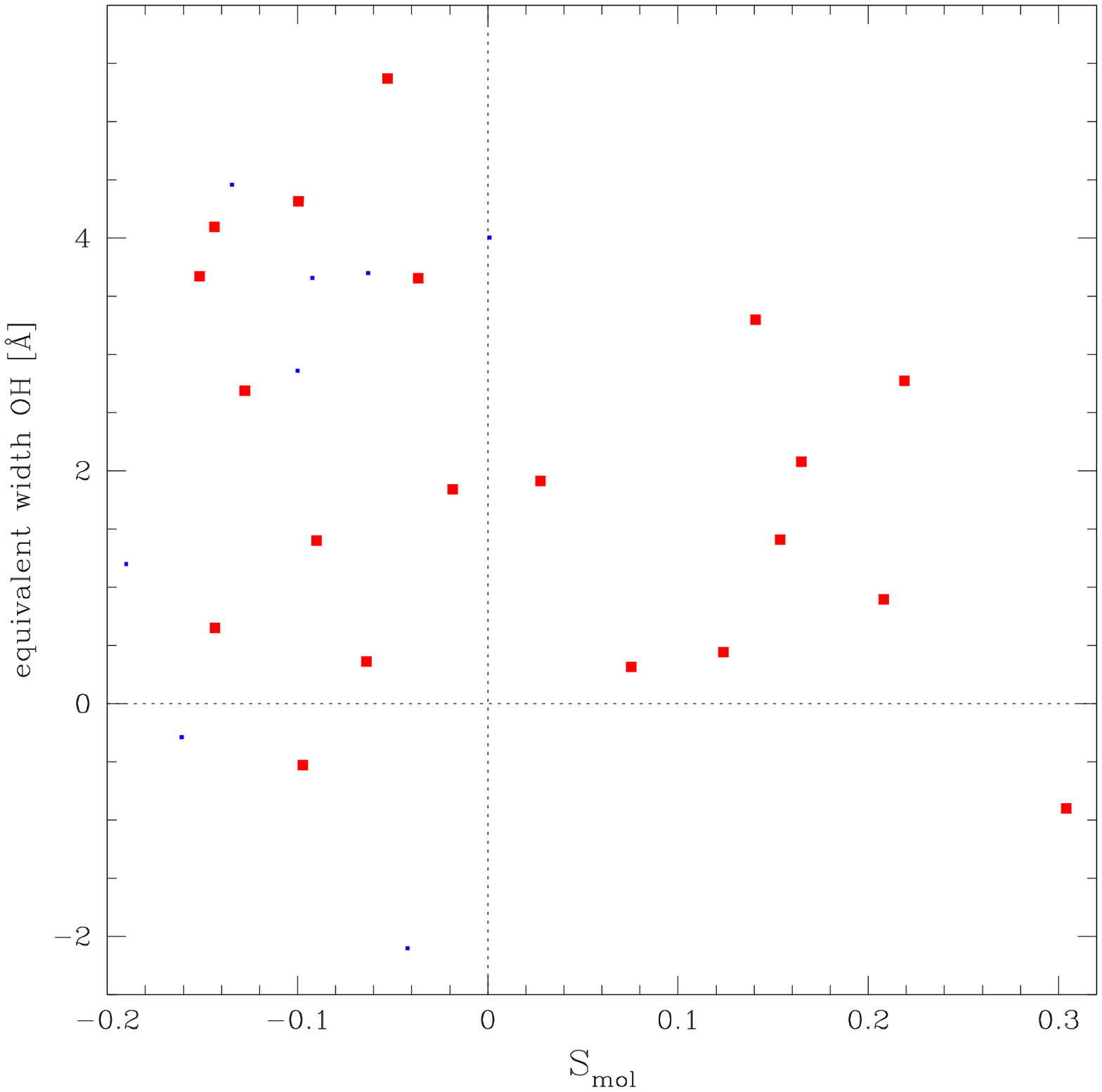,width=59mm}
}}
\caption[]{Molecular index and equivalent width of the OH lines for O-rich AGB
stars and red supergiants in the SMC and LMC.}
\end{figure*}

Stars that do not show red K$_{\rm s}$--L$^\prime_{\rm NB}$ colours indicative
of circumstellar dust tend to have L-band spectra that are not affected by the
molecular bands, and vice versa (Fig.\ 15, left). The redder stars in the SMC
sample do show the molecular bands. This could be associated with a more
advanced evolutionary state of the dustier stars, when they are cooler and
possibly stronger pulsation extends the molecular atmosphere. Indeed, all SMC
stars with spectral types later than M2 show the molecular signature in their
spectra (see Table 6). The S-type star in our sample, BFM\,1 does not show the
molecular bands, possibly because its C/O ratio is close to unity and there is
thus little oxygen left after formation of CO to form water (and SiO).

In contrast to some of the SMC stars, the stars in our LMC comparison sample
(Matsuura et al.\ 2005; van Loon et al.\ 2006) do not show any effect of the
molecular bands on their L-band spectra (Fig.\ 15, left). This could be a
selection bias. However, the M9-type AGB star IRAS\,05128$-$6455 (van Loon et
al.\ 2005a) is very cold and would thus be expected to show strong molecular
bands. Possibly the dust continuum emission veils these bands; dust emission
is certainly very prominent in the RSG OH/IR star IRAS\,05280$-$6910
(NGC\,1984\,IR1; van Loon et al.\ 2005b). It is not obvious why SMC stars with
similar K$_{\rm s}$--L$^\prime_{\rm NB}$ colours should show the spectral
curvature associated with strong molecular bands.

%
%
\begin{figure}
\centerline{\psfig{figure=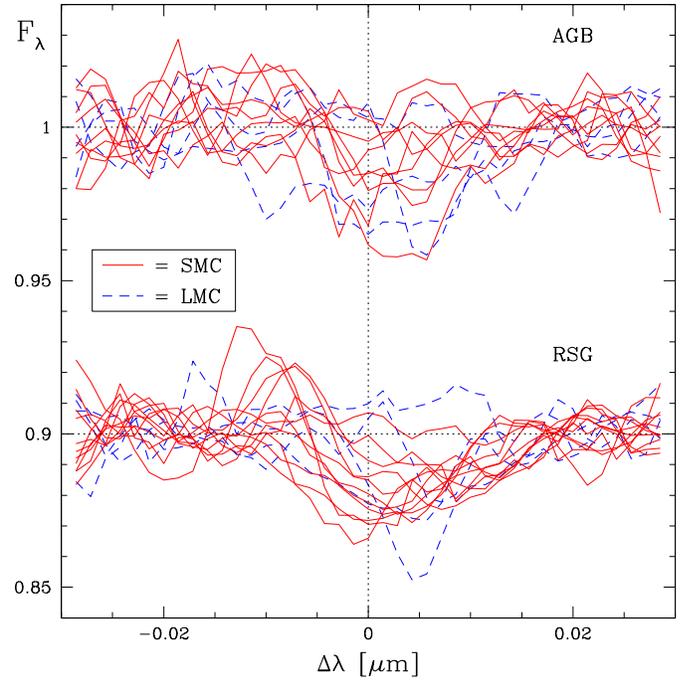,width=88mm}}
\caption[]{Averages of six OH lines (see text), for the O-rich AGB stars and
(offset) for the red supergiants. Both groups of stars show a similar range in
OH absorption. This suggests the presence of OH absorption is not related to
luminosity.}
\end{figure}

Series of equally-spaced lines of hydroxyl (OH) cover parts of the L-band.
These lines are narrow and weak, and hard to see in our spectra. To enhance
our ability to detect OH, we shift-average the spectra such that six of the
stronger, isolated OH lines coincide: OH 1--0 P13--16 and OH 2--1 P14 \& P15
at wavelengths of 3.345, 3.400, 3.455, 3.515, 3.570 and 3.640 $\mu$m,
respectively (Wallace \& Hinkle 2002). As a result, OH is seen in many stars
in our sample, with depths ranging between 0 and $\sim3$\% of the local
(pseudo-)continuum (Fig.\ 16). In this respect, there is no difference between
the O-rich AGB stars and the RSGs, suggesting that luminosity does not play an
important r\^ole in determining the OH abundance. This confirms the findings
by Wallace \& Hinkle (2002) based on Galactic giants and supergiants. The
S-type star, BFM\,1 shows no sign of OH in its spectrum.

The O-rich AGB stars and RSGs in the LMC show a similar range in OH strength
(Fig.\ 16, dashed), from absent in IRAS\,5128$-$6455 and IRAS\,05280$-$6910
--- which have the reddest continua of the O-rich stars in the LMC sample ---
to a depth of $\sim5$\% in the M4-type RSG HV\,2532. The equivalent width of
the average of these OH lines diminishes as the stars become redder (Fig.\ 15,
middle), as expected if the dust continuum veils the OH lines. The less dusty
stars in the LMC show on average stronger OH absorption than SMC stars,
possibly because of the higher oxygen abundance or smaller fraction of oxygen
locked up into CO (or both).

The first overtone rotational-vibrational band of SiO is seen in some stars.
The entire band may be in emission, and is then relatively easy to detect in
low-resolution spectra, e.g., HV\,2084 (Fig.\ 6). In a close-up view of this
spectral region, the band heads can sometimes be seen in absorption, but this
is quite difficult at our signal-to-noise level and low spectral resolution.
Examples are 2MASS\,J00513146$-$7310513 and PMMR\,116, and the absorption
components might also be seen superimposed on the emission in HV\,1375 (Fig.\
17). We list possible SiO detections in Table 6.

%
%
\begin{figure}
\centerline{\psfig{figure=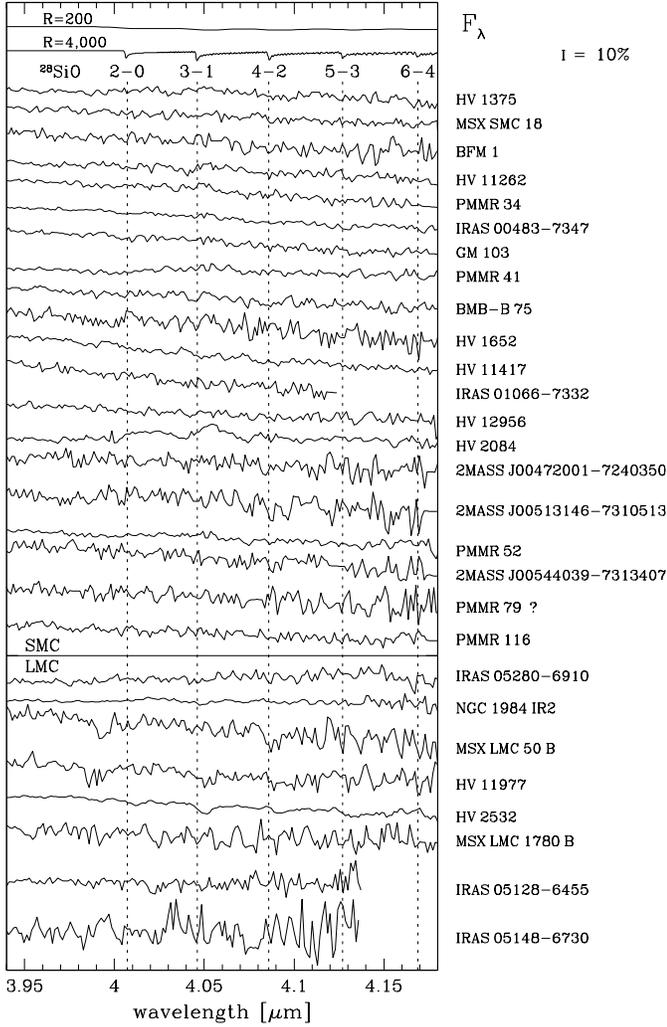,width=88mm}}
\caption[]{The SiO first overtone bands of the O-rich AGB stars and red
supergiants in the SMC and LMC. For comparison, a synthetic SiO spectrum for
$T=2,000$ K is displayed at the top, for two different spectral resolving
powers. The vertical dotted lines indicate the positions of the SiO
bandheads.}
\end{figure}

Matsuura et al.\ (2005) show with medium-resolution spectra that luminous AGB
stars and RSGs in the LMC show hardly any trace of SiO. Veiling by dust and/or
chromospheric continuum emission, and/or band reversal due to pulsational
shocks may have made the SiO bands appear weaker in their small sample (an
example of filling-in of the absorption by diffuse emission is given by Sloan
\& Price (1998) for $\alpha$\,Herculis). Indeed, these LMC stars resemble
luminous, strongly pulsating stars in the Solar Neighbourhood (Rinsland \&
Wing 1982). Our larger compilation of LMC spectra do include some examples of
SiO bands, e.g., HV\,2532 (Fig.\ 17). Targets show the band more often in
emission, whilst serendipitous spectra show it more often in absorption,
likely because the latter tend to have lower luminosity, weaker pulsation and
less continuum veiling. Still, Galactic stars regularly show depths in excess
of 10\% of the continuum, whereas few, if any, of the SMC or LMC stars show
the bands at such level. We thus confirm the tentative conclusion reached by
Matsuura et al.\ (2005), and the firmer conclusion on the basis of the SiO
fundamental band at 8 $\mu$m in Spitzer spectra by Sloan et al.\ (2008), that
SiO is weaker at lower metallicity. The reason is likely the lower silicon
abundance.

HV\,1652 is peculiar. It shows very strong absorption around 2.9 $\mu$m (Fig.\
6), which may be attributed to the 2.7-$\mu$m water vapour band (Merrill \&
Stein 1976). The depression between 3.4 and 3.7 $\mu$m may be due to OH, or
TiO (Schwenke 1998). There is also a hint of narrow SiO lines appearing in
emission (Fig.\ 17). The spectral type of this RSG is only M0.5, and its
K$_{\rm s}$--L$^\prime_{\rm NB}$ is not red at all. It seems to have an
extended molecular atmosphere but not to produce much dust. This might be due
to a lack of nucleation seeds which rely on secondary metallic elements such
as titanium, but if the 3.4--3.7 $\mu$m depression is caused by TiO this would
not appear to be the case.

\section{Discussion}

We first discuss the formation of molecules and dust in carbon- and
oxygen-rich evolved stars in the Magellanic Clouds, before discussing their
impact on the ISM.

\subsection{Molecules and dust in carbon stars}

In the Magellanic Clouds, AGB stars descending from stars that were born
$\sim200$ Myr to a few Gyr ago become carbon stars (Marigo, Girardi \& Chiosi
1996) before they start producing any significant amounts of dust (van Loon et
al.\ 1999b). Near-IR colours are redder, and veiling of the molecular
absorption bands is stronger, in LMC carbon stars compared to SMC carbon
stars. This implies that LMC carbon stars are dustier. The molecular bands are
equally strong in the SMC though, which suggests that the connection between
the molecular atmosphere and the dust envelope is not immediate.

Some SMC carbon stars show as strong absorption by high-elevation acetylene
gas as carbon stars in the LMC. It thus appears that dust forms less
effectively in the SMC from an otherwise similar molecular resource. The
reason for this could be the requirement for nucleation seeds to be present,
which are likely to contain titanium and other secondary elements not produced
by the AGB star itself (e.g., Ferrarotti \& Gail 2006). If less molecular
material is locked in dust, it is not surprising to find relatively large
amounts of remaining molecular gas at high elevation. On the whole, though,
the broader absorption bands in the SMC compared to the LMC suggest that more
of the molecular gas in SMC carbon stars is found close to the surface of the
star (Fig.\ 9).

That the molecular atmospheres do not appear to be {\it more} massive in the
SMC compared to the LMC is in contradiction to the expectations from a simple
scaling with the C/O ratio (Matsuura et al.\ 2005). Either the free carbon
abundance in the SMC does not exceed that in the LMC, or the column density of
molecular gas is limited by other factors such as the temperature and density
profile throughout the pulsating atmospheric layers (Mattsson et al.\ 2008).

Still, it has been suggested that a higher C/O ratio in metal-poor carbon
stars might lead to more prolific dust production in these stars, which would
have important implications for the dust injection rate into the ISM (Matsuura
et al.\ 2005). However, we have just seen that metal-poor carbon stars do not
seem to have dustier winds. If they did, then the dust-driven wind scenario
(van Loon 2000) predicts that their winds should be faster by $\sqrt{\psi}$,
where $\psi$ is the dust-to-gas ratio. For the same total mass-loss rate, the
optical depth of the wind would also increase by the same factor. But the
winds of metal-poor carbon stars are not observed to be optically thicker (see
also Fig.\ 12). To reconcile this with a higher dust content, the mass-loss
rates of metal-poor carbon stars must then be {\it lower}, which would then be
difficult to explain. That not more dust is formed in metal-poor carbon stars
is also consistent with the very low wind speed measured in Galactic halo
carbon stars (Groenewegen, Oudmaijer \& Ludwig 1997; Mauron 2008), and with
the lack of dust seen in Planetary Nebulae in the Magellanic Clouds
(Stanghellini et al.\ 2007).

This all favours the notion of similar total mass-loss rates, via less dusty
winds, of carbon stars in the metal-poor SMC compared to the metal-richer LMC
(van Loon 2000, 2006). The mass-loss rate must be determined by inner boundary
conditions such as pulsation --- which appears to reach similar strength in
both Magellanic Clouds (van Loon 2002; Groenewegen et al.\ 2007) --- whilst
the dust production must be determined by secondary elements such as titanium.

\subsection{Molecules and dust in massive AGB stars and in red supergiants}

The most massive AGB stars and the red supergiants are generally oxygen-rich.
In the SMC, the coolest of these stars show impressively strong molecular
absorption. It is difficult to understand why the molecular abundance should
be so high in these metal-poor stars, in particular as it is also seen in some
of the RSGs --- which are not known to show any surface abundance enhancements
of either oxygen or carbon. The bulk of the molecular mass may be closer to
the surface of the star than in metal-rich stars. This would increase the
column density (less dilution) as well as the excitation temperature, both of
which would result in a higher rate in the rotational-vibrational transitions,
and thus stronger absorption (see Yamamura et al.\ 1999). It is also possible
that the water continuum absorption is optically thick more often in the LMC
stars, in which case the curvature is suppressed. This ``self-veiling'', in
combination with that by the dust continuum, could explain the absence of
curvature and OH lines in the most extreme LMC objects (Fig.\ 15).

The SiO, OH and water seen in the L-band spectrum arise from regions well
within a stellar radius from the stellar photospheric continuum (Tsuji et al.\
1997). The best prospects for detecting these molecules further out are
through the maser lines of SiO at 43 GHz (1--0) and 86 GHz (2--1), water at 22
GHz, and its dissociation product OH at 1612 MHz; these can be stronger than
the thermalised rotational transitions of CO at 115 GHz (1--0) and 230 GHz
(2--1). However, to date no such detections have been made in the SMC, and few
in the LMC (Wood et al.\ 1992; van Loon et al.\ 2001; Marshall et al.\ 2004).
This is likely due to the lower molecular abundances as a result of the lower
abundances of silicon and oxygen in the Magellanic Clouds, and the lower dust
content and hence fainter IR emission and lower pumping rate of the population
inversion that drives the stimulated emission in the maser transition.

The fact that SiO is sometimes seen in emission in the SMC stars implies that
pulsation can be strong in these cool stars despite their low metallicity. It
is therefore conceivable that similar total amounts of gas are deposited in
the region above the stellar photosphere where dust can form. As SiO forms the
basis for the formation of the silicate dust that typically dominates the dust
grain population in well-developed winds from AGB stars and RSGs, a lower SiO
molecular abundance in SMC stars is likely to hamper the growth of dust grains
in their envelopes. Also, the dust formation is likely to rely on a nucleation
seed containing titanium or another secondary element --- Nittler et al.\
(2008) argue that Al$_2$O$_3$ may condense before TiO$_2$ --- which is certain
to be less abundant in low-metallicity stars. However, as long as enough dust
forms to be coupled dynamically with the gaseous fluid, the gas present in the
dust formation z\^one may be driven away albeit at a slower speed. This is
indeed confirmed to happen in OH/IR stars in the Magellanic Clouds compared to
the Milky Way (Marshall et al.\ 2004; van Loon 2006).

\subsection{The r\^ole of pulsation in driving mass loss}

%
%
\begin{figure}
\centerline{\psfig{figure=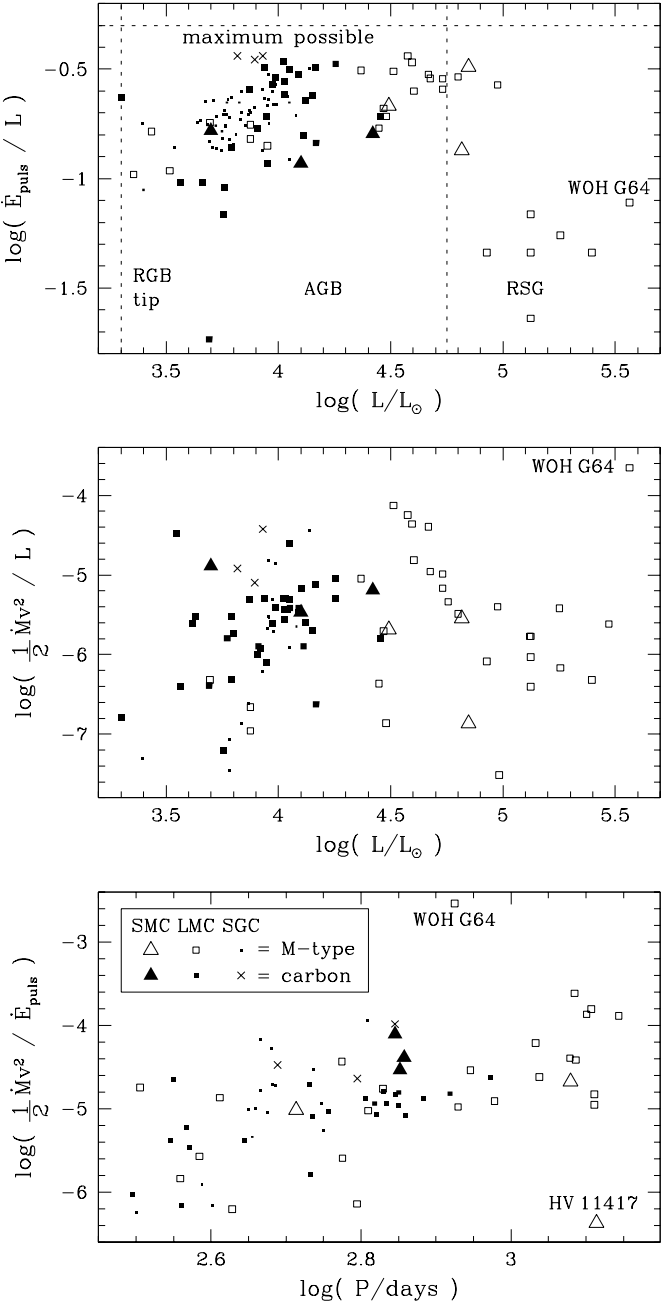,width=88mm}}
\caption[]{The mean energy rate involved in pulsational variability as a
fraction of luminosity, versus luminosity (top), and the kinetic energy in
the ejecta compared to luminosity, versus luminosity (centre) and compared to
pulsational energy rate, versus pulsation period (bottom), for carbon- and
oxygen-rich AGB stars and RSGs in the SMC (triangles), LMC (squares) and South
Galactic Cap (triangles/dots).}
\end{figure}

To quantify the potential of pulsation to drive a wind, van Loon (2002)
introduced the mean energy rate involved in the associated luminosity
modulation (a result of the $\kappa$-mechanism causing the radial pulsation),
taking the K-band amplitude as a proxy for a sinusoidal bolometric light
modulation (cf.\ van Loon et al.\ 2006). This quantity, $\dot{E}_{\rm puls}$,
is mathematically restricted to the range $0-0.5$ L$_\star$, where 0.5
L$_\star$ would correspond to the star ``switching off'' altogether at minimum
light.

We illustrate in Fig.\ 18 the energy contained within the pulsation, wind and
radiation field using mainly data from the LMC (Whitelock et al.\ 2003),
complemented with data from the South Galactic Cap (Whitelock et al.\ 1994);
very limited data is available for the SMC: carbon stars IRAS\,00448$-$7332,
IRAS\,00554$-$7351 and NGC\,419\,IR2 (Groenewegen et al.\ 2007) and M-type
stars IRAS\,00483$-$7347, HV\,11417 and HV\,12956 (Whitelock et al.\ 1989;
Elias, Frogel \& Humphreys 1985) --- all scaled to consistent distances and
dust-driven wind properties (van Loon 2006). There is remarkably little
difference in the trends seen amongst these different populations.

The pulsation of AGB stars appears to saturate just short of the maximum
attainable storage of photons within the partially-ionized regions of the
mantle during expansion (Fig.\ 18, top), whilst the energy output of RSGs is
clearly modulated less strongly. This is reflected in the smaller amplitudes
of the lightcurves of RSGs, but in an absolute sense $\dot{E}_{\rm puls}$ of
RSGs rivals that of AGB stars.

The rate at which the wind carries away kinetic energy is $\dot{E}_{\rm
wind}=\dot{M}v^2/2$, where $\dot{M}$ is the mass-loss rate and $v$ is the wind
speed. Using the data and scaling relations from van Loon et al.\ (1999b,
2005a), the radiation field is found to contain $10^4$ to $10^7$ times more
energy than required to support the wind (Fig.\ 18, centre), independent of
stellar luminosity (cf.\ Judge \& Stencel 1991). This suggests that something
other than radiation determines the mass loss.

The kinetic energy in the wind is tiny ($\sim10^{-5}$ times) compared to the
energy modulated in the pulsating mantle, and a smooth function of pulsation
period, $P$, indistinguishable for M-type and carbon AGB stars (Fig.\ 18,
bottom). WOH\,G064 is an anomaly; its mass-loss rate is over-estimated due to
non-sphericity of the dust envelope (Ohnaka et al.\ 2008). Perhaps slower
pulsations drive a wind more efficiently, if the molecular layers are elevated
more if given more time during a pulsation cycle. Another outlier is
HV\,11417, whose mass-loss rate appears low.

Bowen (1988) found in his pulsation models that the fraction of luminosity
that is dissipated by pulsation shocks is of order a per cent. Less than a per
cent of this appears to be used to perform work on the expanding layers; hence
most of the dissipated energy must leave the system through radiation. Indeed,
modulated H$\alpha$ line emission is ubiquitous in pulsating red giants (e.g.,
McDonald \& van Loon 2007). Bessell et al.\ (1989) found that the temperature
structure generated by Bowen's model does not produce a spectrum that
resembles observed spectra of Mira-type variables, questioning the efficacy of
pulsation in levitating atmospheric layers.

Metal-poor AGB stars need longer to evolve before becoming sufficiently cool
and strong pulsators, delaying the dust-driven wind stage (Bowen \& Willson
1991). This gives the core more time to grow, yielding a more massive white
dwarf remnant. The observed differences between the white dwarf mass functions
in the Magellanic Clouds and the Milky Way are very small (Villaver,
Stanghellini \& Shaw 2004, 2007; Williams 2007), implying that the dust-driven
wind can neither be delayed nor weakened much.

\subsection{Implications for the enrichment and chemical mixing of the
metal-poor interstellar medium}

It is important for understanding the continued star formation and chemical
evolution of the Magellanic Clouds to establish which of the cool giants
contribute most to the replenishment of the Magellanic Clouds' ISM, and how
their material is mixed with the pre-existing ISM.

Carbon stars are plentiful due to a favourable IMF, and lower-metallicity
star-forming galaxies are known to host larger carbon star populations
(Groenewegen 1999). Carbon stars easily produce molecules and dust even at the
low metallicity of the Magellanic Clouds. These molecules will have been
destroyed by the interstellar radiation field (ISRF) before they could take
part in the chemistry within molecular clouds that influences star formation.
But they do supply the ISM with primary carbon. This could push the CO
formation within molecular clouds to be limited by the oxygen abundance, and
the formation of complex molecules to be dominated by carbonaceous chemistry.

Carbon stars in the Magellanic Clouds appear to produce less dust than their
metal-richer siblings, although this is not conclusive until more CO
detections of metal-poor carbon stars become available to measure their wind
speeds and dust-to-gas ratios. Either way, carbon-rich dust has a fairly low
condensation temperature and a high opacity, which means that it captures
photons and disintegrates with much ease. This may explain the ubiquitous
observation of PAHs in photon-dominated regions, both at the interfaces
between H\,{\sc ii} regions and molecular clouds and between the central star
and detached dust envelope within post-AGB objects. Hence the production of
carbon-rich dust may be of little consequence for the ISM. Although the seeds
(TiC$_2$, possibly SiC) may survive and participate in grain growth, the
coatings will be a mixture of oxygen-rich and carbon-rich condensates.

The massive AGB stars and especially the RSGs that dominate the oxygen-rich
dust production stray less far from their natal molecular clouds than carbon
stars do, so the dust and gas they produce will participate in molecular
clouds shielding from the ISRF and chemistry sooner than dust and gas from
carbon stars. Meteoritic data suggest that chemical inhomogeneities exist in
the Galactic ISM (Nittler et al.\ 2008). The slower winds of metal-poor red
giants (Marshall et al.\ 2004; Mauron 2008) result in longer mixing
timescales, also because they carry less momentum and thus stall more easily.
This causes more localised chemical enrichment and continued star formation.
The mixing of the material injected by carbon stars and low-mass oxygen-rich
AGB stars is due in part by ram-pressure stripping of their circumstellar
envelopes (Villaver, Garc\'{\i}a-Segura \& Manchado 2004; Wareing et al.\ 2006)
as they travel through the ISM at speeds of tens of km s$^{-1}$, well in
excess of their wind speeds. Younger, less dynamically relaxed populations are
found within several km s$^{-1}$ from the local ISM velocity (e.g., van Loon
et al.\ 2001).

\section{Summary of conclusions}

We presented 2.9--4.1 $\mu$m spectra, obtained with ISAAC at the ESO/VLT, of a
sample of dusty carbon stars, oxygen-rich AGB stars and red supergiants in the
SMC. Strong absorption bands mainly due to acetylene are seen in the spectra
of carbon stars, whilst many of the oxygen-rich stars too show strong
molecular absorption. OH and/or SiO is detected in some oxygen-rich AGB stars
and red supergiants, with the SiO sometimes seen in emission.

We also presented a spectrum of the post-AGB object MSX\,SMC\,29, which
displays very strong PAH emission. Absorption from aliphatic bonds may have
been detected, which would point at the PAHs being released through
photo-destruction of the grains in the relic AGB envelope. The spectra of two
R\,CrB-type stars show no molecular features but only a dust emission
continuum. We suggest that these may be ``proto-post-AGB'' objects.

The properties of the molecular bands in the SMC sample were compared with
similar data in the LMC. The molecular absorption in SMC carbon stars is as
strong (but not stronger) as that in LMC carbon stars, although there appears
to be more cold molecular gas in the LMC carbon stars. Less conclusive results
are obtained for the oxygen-rich stars. Veiling by dust emission is clearly
more important in the LMC than in the SMC, suggesting that dust formation is
less efficient at lower metallicity. The fact that this is seen also in the
carbon stars suggests that the formation of dust in the molecular atmosphere
relies on an intermediary agent, e.g., a nucleation seed based on a secondary
element such as titanium or silicon. The lower dust content of oxygen-rich
stars may also be caused by lower abundances of secondary elements, in the
nucleation seed or the condensable molecular material.

The pulsation properties of a sample of AGB stars and red supergiants in the
LMC were used to show that the mass-loss rate is likely determined more
critically by the star's ability to levitate the molecular atmosphere through
pulsation, than by the star's ability to drive the elevated molecular layers
away through radiation pressure on the dust (unless dust does not form at
all).

The slower winds of less dusty metal-poor AGB stars and red supergiants were
discussed in the context of the enrichment and mixing of the mass loss with
the ISM. Chemical inhomogeneities and small-scale structure in the dusty ISM
are suggested to be more likely in metal-poor environments.

As a by-product of this programme, we obtained spectra of several Young
Stellar Objects. These include the first detections of water ice in such
objects in the SMC.

\begin{acknowledgements}
We wish to thank the astronomers and telescope operators at ESO Paranal for
the excellent support during our observations, and an anonymous referee for
her/his helpful suggestions. MC thanks NASA for supporting his participation
in this work through contract JPL 1320707 with UC Berkeley. This publication
makes use of data products from the Two Micron All Sky Survey, which is a
joint project of the University of Massachusetts and the Infrared Processing
and Analysis Center/California Institute of Technology, funded by the National
Aeronautics and Space Administration and the National Science Foundation. This
research has made use of the SIMBAD database, operated at CDS, Strasbourg,
France. Based on observations made with the European Southern Observatory
telescopes obtained from the ESO/ST-ECF Science Archive Facility.
\end{acknowledgements}

\appendix
\section{Galactic B-type emission-line stars}

%
%
\begin{figure}
\centerline{\psfig{figure=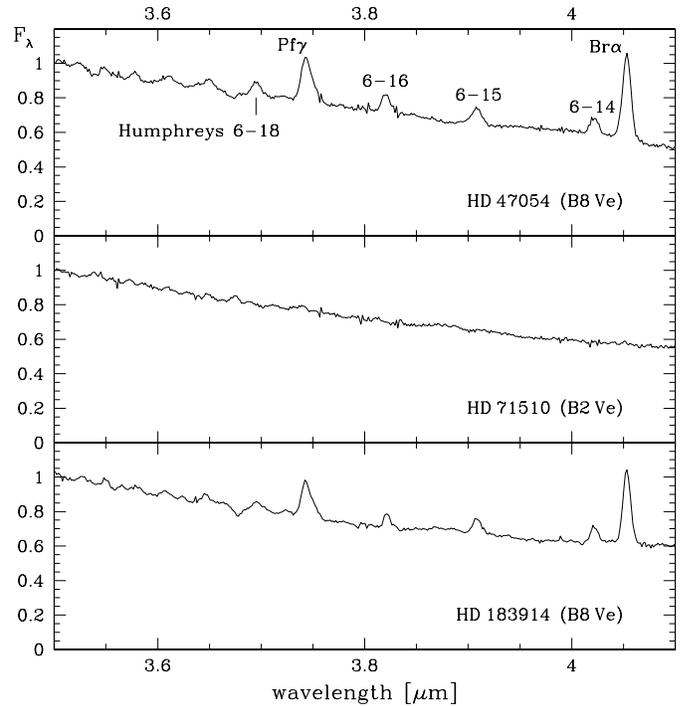,width=88mm}}
\caption[]{Spectra of three Galactic B-type emission-line stars.}
\end{figure}

As part of the telluric standard star calibration plan, three Galactic B-type
stars were observed on 20/21 September 2005 that are now identified in the
Simbad database as emission-line stars: HD\,47054 (HIP\,31583) of type B8\,Ve,
HD\,71510 (HIP\,41296) of type B2\,Ve, and HD\,183914 (HIP\,095951) of type
B8\,Ve.

The spectra of the two B8 stars clearly show hydrogen emission lines from the
Brackett, Pfund and Humphreys series, but the spectrum of the hotter B2 star
is completely featureless (Fig.\ A1).


\begin{thebibliography}{}
\bibitem[]{BernatowiczEtal1991}
Bernatowicz T.J., Amari S., Zinner E.K., Lewis R.S.\ 1991, ApJ, 373, L73
\bibitem[]{BessellEtal1989}
Bessell M.S., Brett J.M., Wood P.R., Scholz M.\ 1989, A\&A, 213, 209
\bibitem[]{BlancoBlancoMccarthy1980}
Blanco V.M., Blanco B.M., McCarthy M.F.\ 1980, ApJ, 242, 938
\bibitem[]{BlancoFrogelMccarthy1981}
Blanco V.M., Frogel J.A., McCarthy M.F.\ 1981, PASP, 93, 532
\bibitem[]{BoothroydSackmann1992}
Boothroyd A.I., Sackmann I.-J.\ 1992, ApJ, 393, L21
\bibitem[]{BoothroydSackmannAhern1993}
Boothroyd A.I., Sackmann I.-J., Ahern S.C.\ 1993, ApJ, 416, 762
\bibitem[]{Bowen1988}
Bowen G.H.\ 1988, ApJ, 329, 299
\bibitem[]{BowenWillson1991}
Bowen G.H., Willson L.A.\ 1991, ApJ, 375, L53
\bibitem[]{CioniEtal2003}
Cioni M.-R.L., et al.\ 2003, A\&A, 406, 51
\bibitem[]{Clayton1996}
Clayton G.C.\ 1996, PASP, 108, 225
\bibitem[]{ClaytonEtal2007}
Clayton G.C., Geballe T.R., Herwig F., Fryer C., Asplund M.\ 2007, ApJ, 662,
120
\bibitem[]{DecinEriksson2007}
Decin L., Eriksson K.\ 2007, A\&A, 472, 1041
\bibitem[]{DefreitaspachecoEtal1998}
de Freitas Pacheco J.A., Barbuy B., Idiart T.\ 1998, A\&A, 332, 19
\bibitem[]{EliasFrogelHumphreys1985}
Elias J.H., Frogel J.A., Humphreys R.M.\ 1985, ApJS, 57, 91
\bibitem[]{FerrarottiGail2006}
Ferrarotti A.S., Gail H.-P.\ 2006, A\&A, 447, 553
\bibitem[]{FilipovicEtal2002}
Filipovi\'c M.D., Bohlsen T., Reid W., Staveley-Smith L., Jones P.A., Nohejl
K., Goldstein G.\ 2002, MNRAS, 335, 1085
\bibitem[]{FleischerGaugerSedlmayr1992}
Fleischer A.J., Gauger A., Sedlmayr E.\ 1992, A\&A, 266, 321
\bibitem[]{FrostEtal1998}
Frost C.A., Cannon R.C., Lattanzio J.C., Wood P.R., Forestini M.\ 1998, A\&A,
332, L17
\bibitem[]{GailSedlmayr1988}
Gail H.-P., Sedlmayr E.\ 1988, A\&A, 206, 153
\bibitem[]{Groenewegen1999}
Groenewegen M.A.T.\ 1999, in: Asymptotic Giant Branch Stars, eds.\ T.\ Le
Bertre, A.\ L\`ebre \& C.\ Waelkens, IAUS 191, p535
\bibitem[]{GroenewegenBlommaert1998}
Groenewegen M.A.T., Blommaert J.A.D.L.\ 1998, A\&A, 332, 25
\bibitem[]{GroenewegenOudmaijerLudwig1997}
Groenewegen M.A.T., Oudmaijer R.D., Ludwig H.-G.\ 1997, MNRAS, 292, 686
\bibitem[]{GroenewegenEtal1995}
Groenewegen M.A.T., Smith C.H., Wood P.R., Omont A., Fujiyoshi T.\ 1995, ApJ,
449, L119
\bibitem[]{GroenewegenEtal2000}
Groenewegen M.A.T., Blommaert J.A.D.L., Cioni M.-R.L., Okumura K., Habing
H.J., Trams N.R., van Loon J.Th.\ 2000, Mem.SAI, 71, 639
\bibitem[]{GroenewegenEtal2007}
Groenewegen M.A.T., et al.\ 2007, MNRAS, 376, 313
\bibitem[]{Hodge1974}
Hodge P.W.\ 1974, PASP, 86, 263
\bibitem[]{HoefnerAndersen2007}
H\"ofner S., Andersen A.C.\ 2007, A\&A, 465, L39
\bibitem[]{HoefnerFeuchtingerDorfi1995}
H\"ofner S., Feuchtinger M.U., Dorfi E.A.\ 1995, A\&A, 297, 815
\bibitem[]{HronEtal1998}
Hron J., Loidl R., H\"ofner S., J{\o}rgensen U.G., Aringer B., Kerschbaum F.\
1998, A\&A, 335, L69
\bibitem[]{JonesMerrill1976}
Jones T.W., Merrill K.M.\ 1976, ApJ, 209, 509
\bibitem[]{JudgeStencel1991}
Judge P.G., Stencel R.E.\ 1991, ApJ, 371, 357
\bibitem[]{KraemerEtal2005}
Kraemer K.E., Sloan G.C., Wood P.R., Price S.D., Egan M.P.\ 2005, ApJ, 631,
L147
\bibitem[]{KraemerEtal2006}
Kraemer K.E., Sloan G.C., Bernard-Salas J., Price S.D., Egan M.P., Wood P.R.\
2006, ApJ, 652, L25
\bibitem[]{LagadecEtal2007}
Lagadec E., et al.\ 2007, MNRAS, 376, 1270
\bibitem[]{LambertEtal2001}
Lambert D.L., Rao N.K., Pandey G., Ivans I.I.\ 2001, ApJ, 555, 925
\bibitem[]{LebertreEtal2005}
Le Bertre T., Tanaka M., Yamamura I., Murakami H., MacConnell D.J.\ 2005,
PASP, 117, 199
\bibitem[]{Lloydevans1980a}
Lloyd Evans T.\ 1980a, MNRAS, 193, 87
\bibitem[]{Lloydevans1980b}
Lloyd Evans T.\ 1980b, MNRAS, 193, 87
\bibitem[]{Lloydevans1983}
Lloyd Evans T.\ 1983, MNRAS, 204, 985
\bibitem[]{LoupEtal1997}
Loup C., Zijlstra A.A., Waters L.B.F.M., Groenewegen M.A.T.\ 1997, A\&AS, 125,
419
\bibitem[]{MarigoGirardiChiosi1996}
Marigo P., Girardi L., Chiosi C.\ 1996, A\&A, 316, L1
\bibitem[]{MarigoGirardiBressan1999}
Marigo P., Girardi L., Bressan A.\ 1999, A\&A, 344, 123
\bibitem[]{MarshallEtal2004}
Marshall J.R., van Loon J.Th., Matsuura M., Wood P.R., Zijlstra A.A.,
Whitelock P.A.\ 2004, MNRAS, 355, 1348
\bibitem[]{MatsuuraEtal2005}
Matsuura M., Zijlstra A.A., van Loon J.Th., et al.\ 2005, A\&A, 434, 691
\bibitem[]{MatsuuraEtal2006}
Matsuura et al.\ 2006, MNRAS, 371, 415
\bibitem[]{MattssonEtal2008}
Mattsson L., Wahlin R., H\"ofner S., Eriksson K.\ 2008, A\&A, 484, L5
\bibitem[]{Mauron2008}
Mauron N., 2008, A\&A, 482, 151
\bibitem[]{McdonaldVanloon2007}
McDonald I., van Loon J.Th.\ 2007, A\&A, 476, 1261
\bibitem[]{MerrillStein1976}
Merrill K.M., Stein W.A.\ 1976, PASP, 88, 285
\bibitem[]{NittlerEtal2008}
Nittler L.R., Alexander C.M.O'D., Gallino R., Hoppe P., Nguyen A.N.,
Stadermann F.J., Zinner E.K.\ 2008, ApJ, in press
\bibitem[]{OhnakaEtal2007}
Ohnaka K., Driebe T., Weigelt G., Wittkowski M.\ 2007, A\&A, 466, 1099
\bibitem[]{OhnakaEtal2008}
Ohnaka K., Driebe T., Hofmann K.-H., Weigelt G., Wittkowski M.\ 2008, A\&A,
484, 371
\bibitem[]{OliveiraEtal2006}
Oliveira J.M., van Loon J.Th., Stanimirovi\'c S., Zijlstra A.A.\ 2006, MNRAS,
372, 1509
\bibitem[]{OliveiraEtal2008}
Oliveira J.M., et al.\ 2008, in preparation
\bibitem[]{Rebeirotetal1993}
Rebeirot E., Azzopardi M., Westerlund B.E.\ 1993, A\&AS, 97, 603
\bibitem[]{RinslandWing1982}
Rinsland C.P., Wing R.F.\ 1982, ApJ, 262, 201
\bibitem[]{Schwenke1998}
Schwenke D.W.\ 1998, in: Chemistry and Physics of Molecules and Grains in
Space. Faraday Discussions No. 109. The Faraday Division of the Royal Society
of Chemistry, London, p321
\bibitem[]{SkrutskieEtal2006}
Skrutskie M.F., et al.\ 2006, AJ, 131, 1163
\bibitem[]{SloanPrice1998}
Sloan G.C., Price S.D.\ 1998, ApJS, 119, 141
\bibitem[]{SloanEtal2006}
Sloan G.C., Kraemer K.E., Matsuura M., Wood P.R., Price S.D., Egan M.P.\ 2006,
ApJ, 645, 1118
\bibitem[]{SloanEtal2007}
Sloan G.C., et al.\ 2007, ApJ, 664, 1144
\bibitem[]{SloanEtal2008}
Sloan G.C., et al.\ 2008, ApJ, submitted
\bibitem[]{SmithSellgrenTokunaga1988}
Smith R.G., Sellgren K., Tokunaga A.T.\ 1988, ApJ, 334, 209
\bibitem[]{SmithEtal1995}
Smith V.V., Plez B., Lambert D.L., Lubowich D.A.\ 1995, ApJ, 441, 735
\bibitem[]{StanghelliniEtal2007}
Stanghellini L., et al.\ 2007, ApJ, 671, 1669
\bibitem[]{TanabeEtal1997}
Tanab\'e T., et al.\ 1997, Nature, 385, 509
\bibitem[]{TanabeEtal1999}
Tanab\'e T., Nishida S., Nakada Y., Onaka T., Glass I.S., Sauvage M.\ 1999,
IAUS, 191, 573
\bibitem[]{TanabeEtal2004}
Tanab\'e T., Ku\v{c}inskas A., Nakada Y., Onaka T., Sauvage M.\ 2004, ApJS,
155, 401
\bibitem[]{TramsEtal1999}
Trams N.R., et al.\ 1999, A\&A 346, 843
\bibitem[]{TsujiEtal1997}
Tsuji T., Ohnaka K., Aoki W., Yamamura I.\ 1997, A\&A, 320, L1
\bibitem[]{Vanloon2000}
van Loon J.Th.\ 2000, A\&A 354, 125
\bibitem[]{Vanloon2002}
van Loon J.Th.\ 2002, in: Radial and Nonradial Pulsations as Probes of Stellar
Physics, eds.\ C.\ Aerts, T.R.\ Bedding \& J.\ Christensen-Dalsgaard, ASPC
259, p548
\bibitem[]{Vanloon2006}
van Loon J.Th.\ 2006, in: Stellar Evolution at Low Metallicity: Mass Loss,
Explosions, Cosmology, eds.\ H.J.G.L.M.\ Lamers, N.\ Langer, T.\ Nugis \& K.\
Annuk, ASPC 353, p211
\bibitem[]{VanloonZijlstraGroenewegen1999}
van Loon J.Th., Zijlstra A.A., Groenewegen M.A.T.\ 1999a, A\&A, 346, 805
\bibitem[]{VanloonMarshallZijlstra2005}
van Loon J.Th., Marshall J.R., Zijlstra A.A.\ 2005b, A\&A, 442, 597
\bibitem[]{VanloonEtal1999b}
van Loon J.Th., Groenewegen M.A.T., de Koter A., Trams N.R., Waters L.B.F.M.,
Zijlstra A.A., Whitelock P.A., Loup C.\ 1999b, A\&A 351, 559
\bibitem[]{VanloonEtal2001}
van Loon J.Th., Zijlstra A.A., Bujarrabal V., Nyman L.-\AA.\ 2001, A\&A, 368,
950
\bibitem[]{VanloonEtal2005a}
van Loon J.Th., Cioni M.-R.L., Zijlstra A.A., Loup C.\ 2005a, A\&A, 438, 273
\bibitem[]{VanloonEtal2005c}
van Loon J.Th., et al.\ 2005c, MNRAS, 364, L71
\bibitem[]{VanloonEtal2006}
van Loon J.Th., Marshall J.R., Cohen M., Matsuura M., Wood P.R., Yamamura I.,
Zijlstra A.A.\ 2006, A\&A, 447, 971
\bibitem[]{VillaverGarciaseguraManchado2004}
Villaver E., Garc\'{\i}-Segura G., Manchado A.\ 2004, RMxAC, 22, 140
\bibitem[]{VillaverStanghelliniShaw2004}
Villaver E., Stanghellini L., Shaw R.A.\ 2004, ApJ, 614, 716
\bibitem[]{VillaverStanghelliniShaw2007}
Villaver E., Stanghellini L., Shaw R.A.\ 2007, ApJ, 656, 831
\bibitem[]{VollmerEtal2006}
Vollmer C., Hoppe P., Brenker F., Palme H.\ 2006, in: $37^{\rm th}$ Annual
Lunar and Planetary Science Conference, \#1284
\bibitem[]{Walker1972}
Walker M.F.\ 1972, MNRAS, 159, 379
\bibitem[]{WallaceHinkle2002}
Wallace L., Hinkle K.\ 2002, AJ, 124, 3393
\bibitem[]{WareingEtal2006}
Wareing C.\ 2006, MNRAS, 372, L63
\bibitem[]{Westerlund1997}
Westerlund B.E.\ 1997, in: The Magellanic Clouds, Camb.\ Astrophys.\ Ser.,
Vol.\ 29
\bibitem[]{WhitelockEtal1989}
Whitelock P.A., Feast M.W., Menzies J.W., Catchpole R.M.\ 1989, MNRAS, 238,
769
\bibitem[]{WhitelockEtal1994}
Whitelock P., Menzies J., Feast M., Marang F., Carter B., Roberts G.,
Catchpole R., Chapman J.\ 1994, MNRAS, 267, 711
\bibitem[]{Williams2007}
Williams K.A.\ 2007, in: $15^{\rm th}$ European Workshop on White Dwarfs,
eds.\ R. Napiwotzki \& M. Burleigh, ASPC 372, p85
\bibitem[]{Woitke2006}
Woitke P.\ 2006, A\&A, 460, L9
\bibitem[]{WoodEtal1992}
Wood P.R., Whiteoak J.B., Hughes S.M.G., Bessell M.S., Gardner F.F., Hyland
A.R.\ 1992, ApJ, 397, 552
\bibitem[]{YamamuraDejongCami1999}
Yamamura I., de Jong T., Cami J.\ 1999, A\&A, 348, L55
\bibitem[]{ZijlstraEtal2006}
Zijlstra A.A., et al.\ 2006, MNRAS, 370, 1961
\end{thebibliography}
\end{document}